\documentclass{aastex}
\usepackage{emulateapj5}
\usepackage{onecolfloat5}
\usepackage{apjfonts}
\usepackage{epsfig}

\def\myputfigure#1#2#3#4#5%
{\vskip#5pt\makebox[0pt]{\hskip#2in
\includegraphics[width=#3\textwidth]{#1}}\vskip#4pt\hfill}

\newcommand\lsim{\mathrel{\rlap{\lower4pt\hbox{\hskip1pt$\sim$}}
        \raise1pt\hbox{$<$}}}
\newcommand\gsim{\mathrel{\rlap{\lower4pt\hbox{\hskip1pt$\sim$}}
        \raise1pt\hbox{$>$}}}

\newcommand{\chisq}{\chi^2}
\newcommand{\nf}{x_{H}}

\newcommand{\strom}{Str\"omgren sphere}
\newcommand{\stromspace}{Str\"omgren sphere~}
\newcommand{\lya}{Ly$\alpha~$}
\newcommand{\lyb}{Ly$\beta~$}
\newcommand{\taudamp}{\tau_{D}}
\newcommand{\taures}{\tau_{R}}
\newcommand{\lobs}{\lambda_{\rm obs}}
\newcommand{\zsource}{z_{s}}

\newcommand{\zstart}{z_{\rm HII}}
\newcommand{\zend}{z_{\rm end}}
\newcommand{\INPUT}{F}

\begin{document}
\twocolumn[
\submitted{To appear in ApJ, vol. 613, 20 September 2004}
\title{Probing the Reionization History Using the Spectra of High-Redshift Sources}

\author{Andrei Mesinger, Zolt\'{a}n Haiman}
\affil{Department of Astronomy, Columbia University, 550 West 120th Street, New York, NY 10027}

\and
\vspace{-0.4cm}
\author{Renyue Cen}
\affil{Department of Astrophysical Sciences, Princeton University, Peyton Hall, Ivy Lane, Princeton, NJ 08544}
\vspace{+0.4cm}

\begin{abstract}
We quantify and discuss the footprints of neutral hydrogen in the
intergalactic medium (IGM) on the spectra of high--redshift (z $\sim$
6) sources, using mock spectra generated from hydrodynamical
simulations of the IGM. We show that it should be possible to extract
relevant parameters, including the mean neutral fraction in the IGM,
and the radius of the local cosmological Str\"omgren region, from the
flux distribution in the observed spectra of distant sources.  We
focus on quasars, but a similar analysis is applicable to galaxies and
gamma ray burst (GRB) afterglows. We explicitly include uncertainties
in the spectral shape of the assumed source template near the Lyman
$\alpha$ line.  Our results suggest that a mean neutral hydrogen
fraction, $\nf$ of unity can be statistically distinguished from $\nf
\approx 10^{-2}$, by combining the spectra of tens of bright ($M
\approx -27$) quasars.  Alternatively, the same distinction can be
achieved using the spectra of several hundred sources that are $\sim
100$ times fainter.  Furthermore, if the radius of the \strom\
can be independently constrained to within $\sim$ 10\%, this 
distinction can be achieved using a single source.
The information derived from such spectra will help in settling the
current debate as to what extent the universe was reionized at
redshifts near z $\sim$ 6.\\ \\
\end{abstract}]

\section{Introduction}

The epoch of cosmological reionization is a significant milestone in
the history of structure formation.  Despite recent observational
break--throughs, the details of the reionization history remain poorly
determined.  The Sloan Digital Sky Survey (SDSS) has detected large
regions with no observable flux in the spectra of several z $\sim$ 6
quasars \citep{becker01, fan03, white03}.  The presence of these
Gunn-Peterson (GP) troughs set a lower limit on the volume weighted
hydrogen neutral fraction of $\nf \gtrsim 10^{-3}$ \citep{fan02}.
This strong limit implies a rapid evolution in the ionizing background
from $z=5.5$ to $z\sim 6$ \citep{cm02, fan02, lidz02, pent02}, and
suggests that we are witnessing the end of the reionization epoch,
with the IGM becoming close to fully neutral at $z\sim 7$ (but see
Songaila \& Cowie 2002 for a different conclusion).  On the other
hand, recent results from the Wilkinson Microwave Anisotropy Probe
(WMAP) have uncovered evidence for a large optical depth to electron
scattering, $\tau_{e} \sim 0.17 \pm 0.04$ \citep{bennett03} in the
cosmic microwave background anisotropies.  Assuming a step-function
model for the reionization history, this result would indicate that
reionization began at z $\sim$ 17 $\pm$ 4 \citep{kogut03, spergel03}.
Although physically motivated ``double'' reionization scenarios,
proposed prior to the WMAP result \citep{cen03a,wl03a} are consistent 
with the combined observations, the details of the reionization process 
remain far from being clear.

Numerous recent theoretical works have addressed resolutions of the
apparent discrepancy (see, e.g., Haiman 2003 for a review).
Successful models incorporate various feedback effects, such as that
due to metal--enrichment \citep{cen03b, wl03b} or associated with the
UV radiation produced by the early ionizing sources
\citep{hh03}. Alternatively, the high--redshift ``tail'' of ionization
has been attributed to an early population of X--ray producing black
holes \citep{ro03, madau03} or even to something more exotic, such as
decaying particles \citep{hanhai03}.  These competing reionization
scenarios each predict a different evolution of the neutral fraction beyond
$z\sim 6$. It would therefore be quite beneficial to observationally
determine the values of $\nf$ at the intermediate redshifts of $6
\lesssim z \lesssim 30$.

As a first step towards this goal, differentiating between a neutral
and a mostly ionized universe at redshifts just beyond $z\sim 6$,
would already aid in discriminating between several models
\citep{hh03}.  To do this however, one requires a more detailed
statistic than the presence or absence of a Gunn-Peterson trough.  In
this paper, we utilize the transmitted flux distribution of a
hypothetical $z\sim 6$ source near its Ly$\alpha$ wavelength.  While
most of the flux on the blue side of Ly$\alpha$ is simply extinguished
for a wide range of neutral fractions ($10^{-3}\lsim \nf\lsim1$),
detectable flux can be transmitted close to the line center, at
wavelengths corresponding to a local HII region around the source
(e.g. Cen \& Haiman 2000; Madau \& Rees 2000).  The spectrum on the
blue side, as well as the flux processed by the damping wing of IGM
absorption and transmitted on the red side of the Ly$\alpha$ line
\citep{m-e98}, depends on the hydrogen neutral fraction of the IGM in
which the source is embedded.  As a result, the flux distributions can
be used as a probe of the neutral fraction in the IGM.

There are two immediate apparent difficulties with the above
approach. First, it requires an estimate of the intrinsic spectrum of
the source.  Second, it requires an ab--initio model of the density
(and velocity) fields surrounding the source, which will influence the
transmitted flux distribution.  Since both of these quantities are
impossible to predict accurately from first principles for any
specific source and line of sight, a single spectrum is unlikely to
provide tight constraints on the neutral fraction.  However, the
hydrogen neutral fraction can still be inferred statistically, by
studying the spectra of a sample of sources.  {\it The purpose of this
paper is to quantify the accuracy to which $\nf$ can be determined in
a future sample of high redshift sources, taking the above
complications into account}.  We use a hydrodynamical simulation to
generate mock spectra of sources for different assumed values of the
hydrogen neutral fraction of the IGM, and quantify the statistical
confidence to which these transmitted spectra can be distinguished
from each other.

The rest of this paper is organized as follows.  In
\S~\ref{sec:signatures}, we discuss the basic signatures of neutral
hydrogen in the IGM that should be imprinted on the spectrum of a
background source.  In \S~\ref{sec:sims}, we describe the method we
use to produce mock spectra.  In \S~\ref{sec:interpret}, we present the
statistical comparison between the various mock spectra, and assess
the accuracy with which the input neutral fraction can be recovered in
each case.  In \S~\ref{sec:issues}, we discuss the relative merits of
using different source types (quasars, galaxies or gamma ray burst
afterglows), and some related issues. In \S~\ref{sec:r_limits}, we explore the benefits of independently constraining the radius of the \strom\.  Finally, we summarize the
implications of this work and offer our conclusions in
\S~\ref{sec:conclude}.

We use redshift, $z$, and the observed wavelength, $\lobs$,
interchangeably throughout this paper as a measure of the distance
away from the source along the line of sight.  These can be related
by: $(1+z) = \lobs$/$\lambda_\alpha$, where $\lambda_\alpha = 1215.67$
\AA\ is the rest-frame wavelength of the \lya line center.  The proper
distance from a source at redshift $\zsource$ to a point at redshift
$z$, is determined by $r$ = $\int_{\zsource}^{z} dz' ~ c
\frac{dt}{dz'}$, where $c(dt/dz')$ is the line element in a given
cosmology, and the comoving distance is $(1+z)$ times the proper
distance.  For reference, in our case a proper distance of 6 Mpc
corresponds to about $\Delta\lambda_{\rm obs}\approx 100$\AA.

Throughout this paper, we assume a standard $\Lambda$CDM cosmology,
with $(\Omega_\Lambda, \Omega_M, \Omega_b, n, \sigma_8,$ $H_0)$ =
(0.71, 0.29, 0.047, 1, 0.85, 70 km/s/Mpc), consistent with the recent
results from WMAP \citep{spergel03}. Unless stated otherwise, all
lengths are quoted in comoving units.

\section{Spectral Signatures of Neutral Hydrogen}
\label{sec:signatures}

In this section, we summarize the spectral signatures of neutral
hydrogen in the IGM.  The spectrum emitted by a source at redshift
$z_s$, $F_0(\lambda)$, will be modified around the Ly$\alpha$
wavelength by absorption by neutral hydrogen atoms along the line of
sight, so that we observe
$F(\lambda)=F_0(\lambda/[1+z])\exp(-\tau[\lambda, z_s])$.  The total
optical depth due to \lya absorption, $\tau$, between an observer at z
= 0 and a source at z = $\zsource$, at an observed wavelength of
$\lobs = \lambda_{s} (1+\zsource)$, is given by:
\begin{equation}
\label{total_tau}
\tau(\lobs) = \int_{0}^{\zsource} dz ~ c \frac{dt}{dz} ~ n_{H}(z) ~ x_{H}(z) ~ \sigma \left( \frac{\lobs}{1+z} \right)
\end{equation}

\noindent where $c(dt/dz)$ is the line element in a given cosmology,
$n_{H}(z)$ is the hydrogen number density at redshift z, $x_{H}(z)$ is
the hydrogen neutral fraction at redshift z, and
$\sigma(\frac{\lobs}{1+z})$ is the Ly$\alpha$ absorption cross section.
Since high-redshift sources sit in their own highly ionized
Str\"omgren spheres\footnote{The assumption of discrete HII regions is
invalid if hard spectra dominate reionization. In this case, the
surface of the HII region can have a width that exceeds the separation
between the ionizing sources \citep{oh01, vgs01}.  There is still a
proximity region around each source, within which they dominate the
cosmic background, which could be used in the analysis below in place
of the Str\"omgren sphere.}, the total optical depth at a given
$\lobs$ can be written as the sum of contributions from inside and
outside the \strom, $\tau = \taures + \taudamp$.  The resonant optical
depth, $\taures$, is given by:

\begin{equation}
\label{res_tau}
\taures(\lobs) = \int_{\zstart}^{\zsource} dz ~ c \frac{dt}{dz} ~  n_{H}(z) ~ x_{H}(z) ~ \sigma \left( \frac{\lobs}{1+z} \right)
\end{equation}

\noindent and the damping wing optical depth for the IGM outside the
Str\"omgren sphere can be obtained from

\begin{equation}
\label{damp_tau}
\taudamp(\lobs) = \int_{z_{end}}^{\zstart} dz ~ c \frac{dt}{dz} ~ n_{H}(z) ~ \nf(z) ~ \sigma \left( \frac{\lobs}{1+z} \right)
\end{equation}

\noindent where $\zstart$ corresponds to the redshift of the edge of
the \strom, and $z_{end}$ denotes the redshift by which HI
absorption is insignificant along the line of sight to the source.
For a smooth IGM, we would simply have $n_{H}(z) = n_{H,0}~(1+z)^3$,
and since the fall-off of the cross section is rapid compared to the
change in $x_{H}(z)$, we can also approximate $\nf \simeq constant$,
for the smooth IGM in eq. (\ref{damp_tau}).  Conversely, the fall-off
of the cross section is slow compared to the small-scale fluctuations
in the IGM, averaging out their contributions to $\taudamp$.  For
these reasons, in this paper we approximate $\nf \simeq constant$, and
assume a smooth IGM.  The value of $z_{end}$ was chosen to be $z_{end}
= \zstart - 0.5$.  As long as we are not looking through another
source's Str\"omgren sphere close to $\zstart$, we find that the exact
value of $z_{end}$ is irrelevant because most of the contribution to
$\taudamp$ comes from within $\Delta z<0.5$ of $\zstart$.  
In particular, a choice of $z_{end} = \zstart - 0.25$ results in
an average change in $\taudamp$ of only $\sim 4\%$ compared to our
fiducial model with $z_{end} = \zstart - 0.5$.
The difference between equations~(\ref{res_tau}) and (\ref{damp_tau}) are:
(1) the limits of integration, (2) the determination of $x_H$, which
assumes ionization equilibrium with the cosmological background flux
$J_{\rm BG}$ outside the Str\"omgren sphere with the approximation of
$\nf \simeq constant$, and equilibrium with the sum of the background
and the local source fluxes, $J_{\rm BG} + J_{\rm s}$ inside the HII
region with a fluctuating $\nf(z)$, and (3) at the relevant
wavelengths where $\tau_R$ is significant, the integral in
equation~(\ref{res_tau}) is dominated by the resonant cross--sections,
whereas at all wavelengths where flux is transmitted, the damping wing
dominates equation~(\ref{damp_tau}).  As a result of this last
property, $\tau_R$ fluctuates strongly with wavelength, reflecting the
local density field, whereas the damping wing contribution is smooth,
since it averages the contributions to the damping wing from a
relatively wide redshift interval.

The two optical depths are represented in Figure~\ref{fig:schematic}
for the case of a bright quasar embedded in a neutral IGM at
$\zsource=6$.  The opacities were obtained from a simulation
(described below) for a typical line of sight towards a source
residing at a cosmological density peak.  Also shown is $R_S$, the
radius of the \strom, corresponding to the region of transmitted flux
blueward of the \lya line center. It is worthwhile to note that there
are different wavelength regions where either the resonant or the
damping wing absorption dominates. These regions can shift according
to the epoch being studied, since $\taudamp$ scales linearly with
$\nf$, and also according to the quasar's luminosity $L$, since
$\taures \varpropto L^{-1}$ (see eq.~[\ref{neutralfraction}]).  For
example, since the damping wing is less dominant close to the line
center for lower $\nf$, details of the gas density and velocity
distribution can be studied in this regime. Shortward of the edge of
the \strom, no flux is observed for $10^{-3}<\nf<1$, since the
attenuation is large (ranging from $\sim
\hspace{0.3cm} e^{-10^6}$ -- $e^{-10^3}$ for the values of the neutral
fraction studied in this paper).

Simple modeling predicts several distinctive signatures that $\nf$
should leave on a spectrum (see also Fig.~\ref{fig:4specs} and
accompanying discussion below).  Most importantly: on average, a
neutral universe would be expected to have a smoother spectrum on the
blue side of the line, due to the larger contribution to the total
optical depth from the IGM, which is a smooth function (see
Fig.~\ref{fig:schematic}).

In addition, the {\it symmetry} of the observed spectrum around the
\lya line center should be affected by $\nf$.  Since $\taures
\rightarrow 0$ on the red side of the \lya line, the observed spectrum
should trace out the emitted spectrum for low values of $\nf$.  
For low values of $\nf$, the red side of the line has negligible
absorption, while the blue side is still affected by resonance
absorption. This makes the observed spectrum highly asymmetric.  In
comparison, in a neutral universe, the presence of a strong damping
wing ($\taudamp$) causes additional strong suppression on the red side
of the line, as well, making the line more symmetric (though overall
weaker).  Note that if we were to model out $\taures$, leaving only
the damping wing contributing to optical depth, a neutral universe
would have a more asymmetric spectrum, because the damping wing would
impose a sharper slope in the transmitted spectrum near the \lya line.
 Furthermore, in a composite spectrum, one expects there to be a
sharp drop in observed flux at the edge of the Str\"omgren sphere if
the universe is mostly ionized.  This occurs since the gradual cusp of
the damping wing (immediately longward of the sharp rise in
$\taudamp$, c.f. Fig.~\ref{fig:schematic}), is sub-dominant if the
universe is mostly ionized, making the transition from
resonance-dominated to damping wing-dominated optical depth more
sudden (see feature at 8390
\AA, in the lower panel of Fig. 3).  On the other hand, if the
universe is neutral, the damping wing cusp dominates at the edge of
the Str\"omgren sphere, and creates a gradual drop-off in the observed
flux (see the upper panel of Fig. 3).  This effect is only observable
if the Str\"omgren sphere is not large enough that $\taures$, whose
mean value scales as the square of the distance from the source,
blocks off the observed spectrum by itself.

\vspace{+1\baselineskip}
\myputfigure{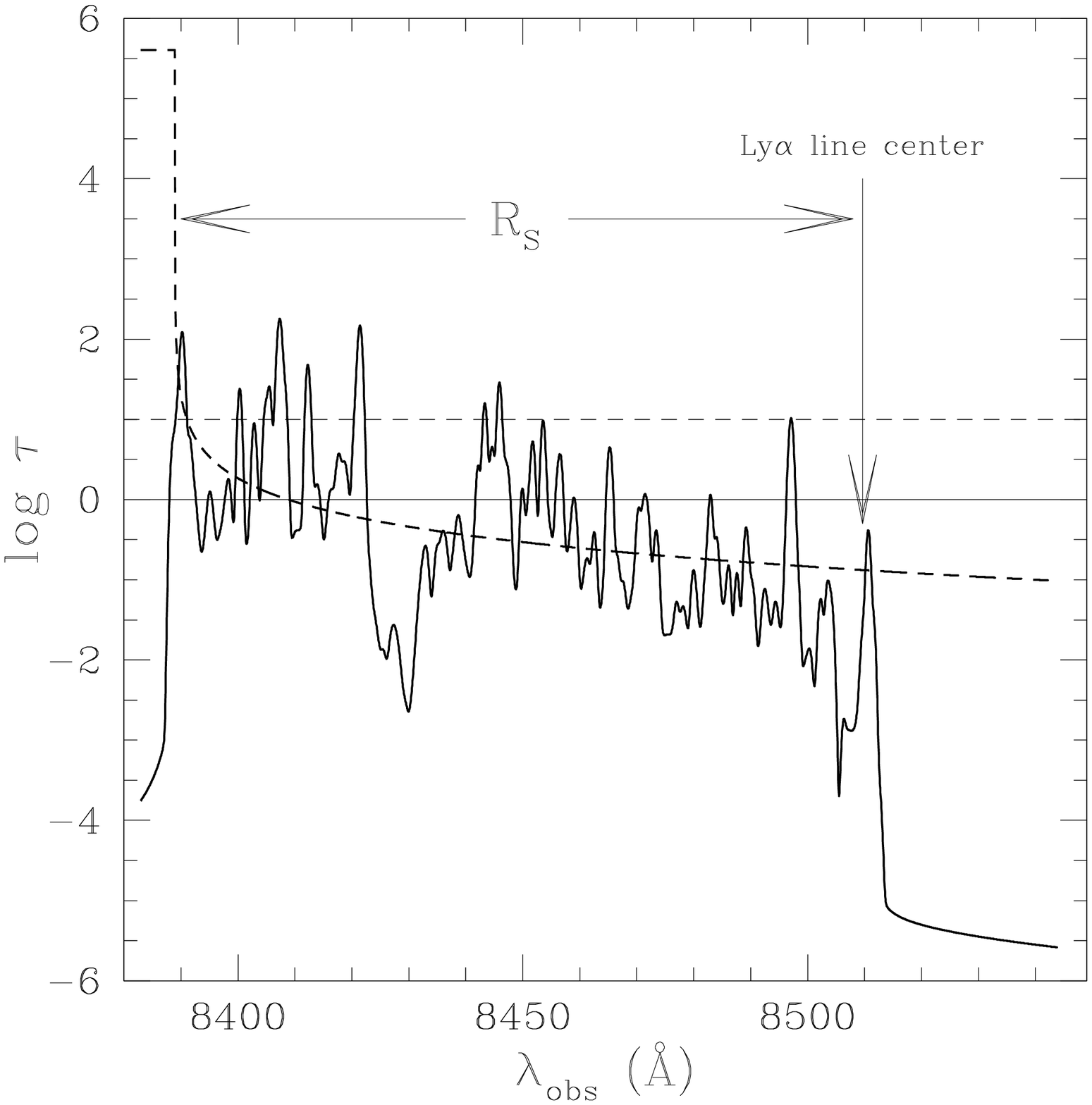}{3.3}{0.5}{.}{0.}
\vspace{-1\baselineskip} \figcaption{Optical depth contributions from
within ($\tau_R$) and from outside ($\tau_D$) the local HII region for
a typical line of sight towards a $\zsource=6$ quasar embedded in a
fully neutral IGM.  The \emph{dashed line} corresponds to $\taudamp$,
and the \emph{solid line} corresponds to $\taures$.  For details
concerning the simulation, see \S~\ref{sec:sims}.  In this example,
the damping wing of the IGM, $\taudamp$, contributes significantly to
the total optical depth at $\lobs \sim 8430$ \AA\ and $\lobs \gtrsim
8470$ \AA.\label{fig:schematic}}
\vspace{+1\baselineskip}

The above effects manifest themselves in all high-redshift sources
(quasars, galaxies, GRB afterglows), and furthermore, they also scale
predictably with the source strength and environment.\footnote{Spectra
of GRB afterglows originating in high redshift galaxies might not have
noticeable $\taures$, since the host galaxies could be too young or
too faint to produce a significant \strom~(Barkana \& Loeb 2004, Lamb
\& Haiman 2003).}  As we will see below, the effects are difficult to
detect or quantify for a single high-redshift spectrum, due to the
uncertainty in the shape of the source's intrinsic emission, and to
the stochastic nature of the density field along a given line of
sight.  However, with many sources such statistics can be usefully
analyzed.

\section{Simulation of Spectra}
\label{sec:sims}

We use a $\Lambda$CDM hydrodynamical simulation box at redshift
$\zsource$ = 6 as the home of our source quasar.  The simulations are
described in detail in (Cen et al. 2004), and we only briefly
summarize the relevant parameters here. The box is 11 $h^{-1}$ Mpc on
each side, with each pixel being about 25.5 $h^{-1}$ kpc.  This scale
resolves the Jeans length in the smooth IGM by more then a factor of
10.  The simulation also includes feedback in the form of realistic
galactic winds.  Stellar particles are treated dynamically as
collisionless particles, except that feedback from star formation is
allowed in three forms: UV ionizing field, supernova kinetic energy,
and metal rich gas, all being proportional to the local star formation
rate.  Supernova energy and metals from aging massive stars are
ejected into the local gas cells where stellar particles are located.
Supernova energy feedback into the IGM is included with an adjustable
efficiency (in terms of rest-mass energy of total formed stars) of
$e_{SN}$, which is normalized to observations (Cen et al. 2004, in
preparation).

We first identified the densest region in the box, as the natural
location of a high--redshift source, such as a quasar. This
corresponds to a pixel with an overdensity of $n/\langle n \rangle
\sim 10^4$ relative to the background, and to the center of a
collapsed dark matter halo with mass, $M_{halo} \approx 2\times
10^{10} M_{\sun}$. Density and velocity information was then extracted
from 92 different lines of sight (LOSs), approximately evenly spaced
in solid angle, originating from this pixel. The step size along each
line of sight (LOS) was taken to be 5.1 $h^{-1}$ kpc, which resolves
the \lya Doppler width by more than a factor of 40.  
The exact value of the step size was chosen somewhat arbitrarily,
and does not influence the results as long as it adequately resolves
the Doppler width.
  At each step, the
density and velocity values were averaged for the neighboring pixels,
and weighted by the distance to the center of the pixels.  We extended
each LOS by the common practice of randomly choosing a LOS through the
box, and stacking the pieces together \citep{cen94}.

The hydrogen neutral fraction inside the \strom, $x_H(z)$, was
calculated at each step in the LOS, and for several values of an
isotropic background flux, $J_{BG}(\nu)$ (in units of ${\rm erg\,
s^{-1}\, cm^{-2}\, Hz^{-1}\, sr^{-1}}$), corresponding to a particular
$\nf$.  Assuming ionization equilibrium, and $x_H(z) \ll 1$ inside the
\strom:

\begin{equation}
\label{neutralfraction}
x_H(z) = \frac{n_H \alpha_B}{\int_{\nu_H}^{\infty} (\frac{L_\nu}{4 \pi r^2} + 4 \pi J_{BG}) \frac{\sigma}{h\nu} d\nu}
\end{equation}

\noindent where $\nu_H$ and $\sigma$ are the ionization frequency and
cross section of hydrogen, respectively, $\alpha_B$ is the hydrogen
recombination coefficient at T = 15,000 K, $r$ is the luminosity
distance between the source and redshift $z$, and $L_\nu$ is the
quasar's intrinsic luminosity in $\rm erg~s^{-1}~Hz^{-1}$.  The
luminosity was taken to be $L_\nu= 2.34 \times 10^{31}
(\nu/\nu_H)^{-1.8} [(1+z)/(1+\zsource)]^{-0.8}$, which results from
redshifting a power-law spectrum with a slope of $\nu^{-1.8}$,
normalized such that the emission rate of ionizing photons per second
is $2 \times 10^{57}$, matching the \citet{elvis94} template spectrum
and the brightness typical of the SDSS quasars.  The background flux,
$J_{BG}(\nu)$, is also assumed to follow a $\nu^{-1.8}$ power-law
spectrum.  Results are insensitive to the shape of the background
flux, since the dominant effect of $J_{BG}(\nu)$ comes from the
damping wing, and that only depends on the value of $\nf$. Since the
quasar's luminosity is dominant for the values of interest inside the
\strom, we find that equation~(\ref{neutralfraction}) can be very well
approximated by $x_H(z) = 4 \pi r^2 n_H \alpha_B /
\int_{\nu_H}^{\infty} (L_\nu \sigma / h \nu) d\nu$.

For several combinations of $\zstart$ (corresponding to a Str\"omgren
sphere radius, $R_S$) and $\nf$, the integrals in (\ref{res_tau}) and
(\ref{damp_tau}) were evaluated for each LOS.  To expedite analysis,
$\taudamp$ was also calculated using the approximation \citep{m-e98}:

\begin{eqnarray}
\label{jordi_tau}
\taudamp &=& 6.43 \times 10^{-9} \nf \left( \frac{\pi e^2 f_{\alpha} n_{H}(z)}{m_e c H(z)} \right)\\
\nonumber &\times& \left[ I\left( \frac{1+\zstart}{1+z}\right) - I\left( \frac{1+\zend}{1+z}\right)  \right]
\end{eqnarray}

\noindent where:

\begin{eqnarray}
\nonumber I(x) &\equiv& \frac{x^{9/2}}{1-x} + \frac{9}{7} x^{7/2} + \frac{9}{5} x^{5/2} + 3x^{3/2} + 9x^{1/2} \\
\nonumber &-& \ln \left| \frac{1+x^{1/2}}{1-x^{1/2}} \right|
\end{eqnarray}
\\
\noindent Equations (\ref{damp_tau}) and (\ref{jordi_tau}) give very
similar results for wavelengths away from resonance (i.e. inside the
\strom), but (\ref{jordi_tau}) is much quicker to compute.

The Voigt profile was used to approximate the Ly$\alpha$ absorption cross
section \citep{rl79}.  We assumed a temperature of 15,000 K for gas
inside the \strom.  Outside the Str\"omgren sphere, for the case of a
mostly ionized universe, an IGM temperature of $T = 10^4$ K was used,
while the temperature in a neutral universe was taken to be $T = 2.73
\times 151 (\frac{1+z}{151} )^2$ K, valid for $z<150$
\citep{peebles93}.  Our results are insensitive to the exact
temperature used.  The Doppler width of the \lya absorption cross
section scales as $\nu_D \propto T^{1/2}$, but the total integrated
area under the cross section is independent of temperature.

Finally, to simulate observations, we had to smooth the raw spectra we
compute.  Physical smoothing due to gas pressure, present in the
simulations on scales of $\sim$ 10 km/s \citep{gnedin00} corresponds
to smoothing on a wavelength range of $\Delta \lobs = \Delta \lambda_s
(1+\zsource)$, or $\sim$ 0.3 \AA\ at $\zsource=6$.  However, current
spectral resolutions achieved for high--redshift quasars (APO, Keck)
are about a factor of three worse than this value. In order to
simulate a realistic spectral resolution of $\sim$ 1 \AA, all
resulting spectra were smoothed over 1 \AA\ bins  (20 steps in a LOS) by averaging
$e^{-\tau}$ over each bin.

\vspace{+1\baselineskip}
\myputfigure{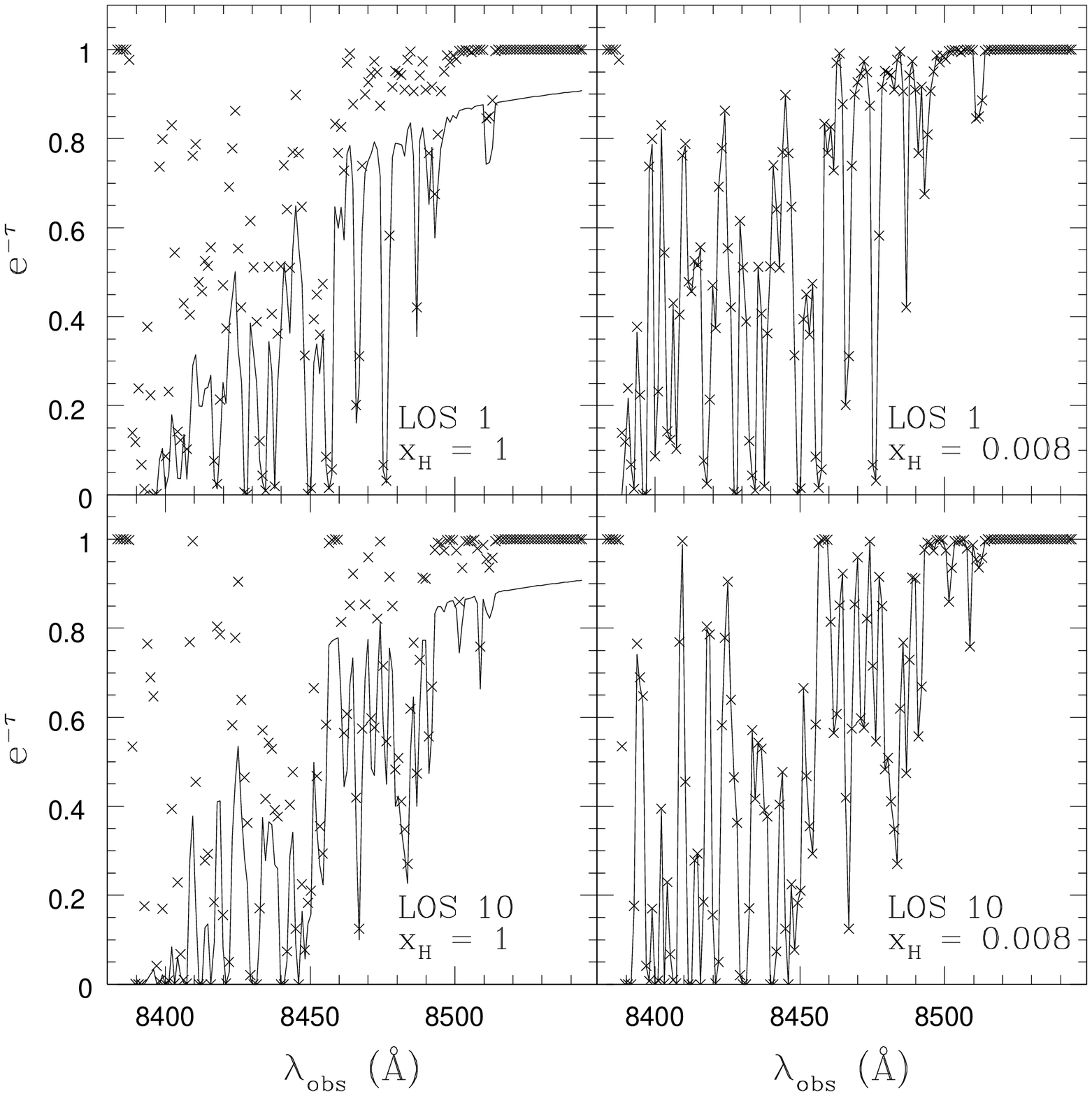}{3.3}{0.5}{.}{0.}  
\vspace{-1\baselineskip} \figcaption{Simulated $e^{-\tau}$ for two
different LOSs (\emph{top and bottom}) in two different $\nf$ regimes:
$\nf = 1$ (\emph{left panels}) and $\nf = 0.008$ (\emph{right
panels}).  All spectra correspond to $R_S = 43 ~ \rm Mpc$.  The solid
line corresponds to $e^{-\tau}$ from equation (\ref{total_tau}) and
the discrete points correspond to $e^{-\taures}$ from equation
(\ref{res_tau}).  The sharp absorption features in the figures are
produced by $\taures$, while the general trend of increasing optical
depth away from the line center is a combination of both $\taures$ and
$\taudamp$.\label{fig:4specs}}
\vspace{+3\baselineskip}

\section{Interpretation of Spectra}
\label{sec:interpret}

For illustrative purposes, in Figure 2 we show four different examples
for the spectrum of a source with $R_S = 43 ~ {\rm Mpc}$.  We show
spectra assuming two different values of $\nf$: 1 or 0.008,
corresponding to $J_{BG}(\nu_H)$ = 0 and $8 \times 10^{-25}$ ${\rm
erg~s^{-1}~cm^{-2}}$ ${\rm Hz^{-1}~sr^{-1}}$, respectively.  This
range for the background flux is approximately what is allowed by the
current analysis of known $z\sim 6$ quasars (e.g. Fan et al. 2003; Cen
\& McDonald 2002).  The value of $R_S$ was chosen because it is
representative of the brightest SDSS quasars, and it corresponds to
the radius of a sphere enclosing $\sim 2 \times 10^{57}$ ionizing
photons per second $\times~10^{15}$ s hydrogen atoms \citep{ch00}.
The value of the source's Str\"omgren sphere radius is left as a free
parameter to be determined by the inversion method (described below).

Also, shown in Figure 2 is the effect of the damping wing in smoothing
out the spectrum.  The flux decrement due to resonance absorption,
$e^{-\taures}$, is shown with discrete points, and the total flux
decrement, $e^{-\taures - \taudamp}$, is represented with a solid line
for two different LOSs (\emph{top and bottom}).  Flux decrements from
a neutral universe (\emph{left panels}) are visibly smoother than
those of a mostly-ionized universe (\emph{right panels}), as
predicted.  The sharp absorption features in the figures are produced
by $\taures$, while the general trend of increasing optical depth away
from the line center is a combination of both $\taures$ and
$\taudamp$.

\vspace{+1\baselineskip}
\myputfigure{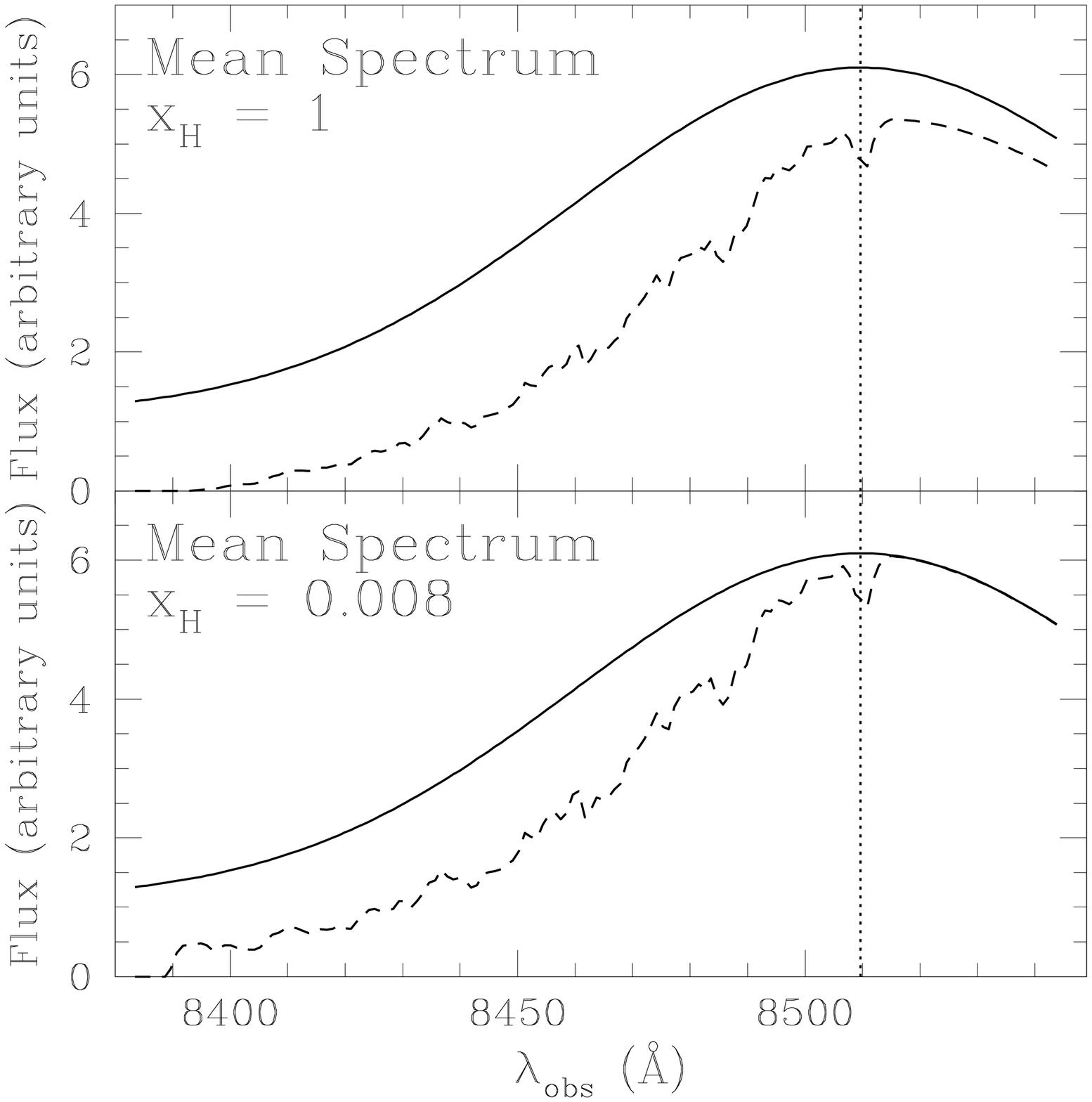}{3.3}{0.5}{.}{0.}  
\vspace{-1\baselineskip} \figcaption{Composite mean spectra from 92
LOSs, for the cases of $\nf = 1$ (\emph{top panel}) and $\nf = 0.008$
(\emph{bottom panel}).  For illustrative purposes, a Gaussian +
continuum emission template is overlaid (\emph{solid line}), and the
resulting mean observed spectrum is shown underneath (\emph{dashed
line}).  The source quasar is at redshift, $\zsource = 6$, and the
\lya line center is at 8509.69 \AA (\emph{dotted line}).  The spectrum shown in the top
panel is measurably smoother than the one shown in the bottom panel.}
\vspace{1.5\baselineskip}

In Figure 3, we show the composite mean spectra from 92 LOSs, for the
two different values of $\nf$: 1 (\emph{top panel}) and 0.008
(\emph{bottom panel}).  For illustrative purposes, a Gaussian +
continuum emission template is overlaid (\emph{solid line}), and the
resulting mean observed spectrum is shown underneath (\emph{dashed
line}).  
The emission template was chosen somewhat arbitrarily, in order
to resemble the typical spectrum of observed quasars (e.g. Vanden Berk
et al. 2001), and is comprised of a flat continuum emission added to a
Gaussian of width 2500 km/s and a peak--to--continuum ratio of $\sim$ 5. 
From Figures 2 and 3, we can see that the composite mean
spectra for these regimes exhibit the traits discussed in \S~\ref{sec:signatures}: 
spectra from a neutral universe are smoother on the blue
side of the line and more symmetric around the line center (at 8509.69
\AA); however, if the $\taures$ contribution is statistically modeled
out, the $\nf = 0.008$ spectra would be more symmetric; also there is
a sharp drop in the composite observed flux at the edge of the
Str\"omgren sphere in the $\nf = 0.008$ universe, that is not present
in the neutral universe composite spectrum.  The latter feature is
only statistically detectable from a large sample of LOSs, and was not
used in our analysis below.  In the analysis below, we ignore the
$\sim$ 5 \AA\ range redward of the edge of the \strom, since in this
region, the exact structure of the transition region from the
\strom~to the IGM contributed significantly to $\taudamp$, and the
transition region can have a large spectrum-to-spectrum variability.

\subsection{Inverting the Observed Spectrum to Find $\nf$}
\label{sec:invert_spect}

We are now presented with an interesting problem: supposing the only
information we have is the observed spectrum, can we extract the
quantities of interest, namely $\nf$ and $R_S$?  For the purposes of
an idealized analysis, we will first assume that we know the {\it
shape} of the intrinsic emission template to infinite precision
(however, we do not assume knowledge of the {\it amplitude} of the
template).  In the next section, we shall relax this restriction. The
other major assumption we implicitly make is that we can accurately
predict the distribution of $\taures$, using the small-scale power
statistics from the simulation box.  This assumption can be checked by
studying more extensive simulations in the future (see discussion in
\S~\ref{sec:issues} below).

Before diving into the details, we first outline the main steps
of the procedure.  We start with a simulated observed spectrum.  Then
we guess values for the radius of the \strom, $R_{S}'$, and the IGM
hydrogen neutral fraction, $\nf'$.  Next we approximate the amplitude
of the source's intrinsic emission, $A'$, implied by the choices of
$R_{S}'$ and $\nf'$, using the red side of the \lya line where
resonance absorption can be neglected.  From the observed spectrum, we
divide out the assumed intrinsic emission ($A'$ $\times$ {\it known
spectral shape}), and the assumed damping wing flux decrement,
$e^{\taudamp(\lobs, R_{S}',\nf')}$, calling the result $S'(\lobs)$.
If our choices of $R_{S}'$ and $\nf'$ were correct, $S'(\lobs)$ should
represent the resonance absorption flux decrement alone.  Hence, we
compare a histogram of the implied resonance optical depths, $-
\ln[S'(\lobs)]$, to the known histogram of resonance optical depth
(obtained from the simulation).  We then repeat this procedure with
different choices of $R_{S}'$ and $\nf'$, finding the ones whose
implied resonance optical depths most closely match the known
histogram.  We shall now elaborate on this procedure below.  

We start with the mock observed flux, of the form: 

\begin{equation}
\INPUT(\lobs) = A ~ e^{-\taures(\lobs) - \taudamp(\lobs, R_{S},\nf)} ~ T(\lobs)
\label{input_flux}
\end{equation}

\noindent where A is the amplitude of the adopted normalized emission
template $T(\lobs)$, and $\taures(\lobs) \equiv \taures(\lobs,
R_{S},\nf)$ also takes into account the small contribution of $J_{BG}$
to $x_H(z)$.  Since we assume we know the shape of the source's
intrinsic emission, for simplicity we shall set $T(\lobs) = 1$.
Actual features in the spectrum will only effect the signal--to-noise
of detection under this optimistic assumption (see discussion below).
Next, we guess values for $R_{S}'$ and $\nf'$.  For these values,
$\taudamp(\lambda_{red}, R_{S}',x_{H}')$ is calculated for a
wavelength, $\lambda_{red}$, on the red side of the line.  We then
estimate the amplitude of the intrinsic emission template, which these
values would imply at $\lambda_{red}$: $A'$ =
$\INPUT(\lambda_{red})$~$e^{\taudamp(\lambda_{red}, R_{S}',x_{H}')}$.
Since for the red side of the line, $\taures \rightarrow 0$, we are
left with the following estimate of the emission amplitude:

\begin{equation}
\label{probed_A}
A' \simeq \langle A ~ e^{\taudamp(\lambda_{red}, R_{S}',\nf') -
\taudamp(\lambda_{red}, R_{S}, \nf)} \rangle
\end{equation}

\noindent where for increased accuracy, $A'$ can be averaged over
several values of $\lambda_{red}$, corresponding to a smooth region in
the red side of the observed spectrum.  Since our simulated spectra
are well-behaved, averaging was not needed, and $A'$ was evaluated at
an arbitrarily chosen $\lambda_{red} = 8514$ \AA.  It should be noted
that, when using real spectra, one has to be careful to chose
$\lambda_{red}$ values that are not near other emission lines.

Next, we divide the input flux by $A'$, and extract an estimate of the
resonant optical depth, $\hat{\taures}(\lobs)$, using the damping wing
contribution, $e^{\taudamp(\lobs, R_{S}',\nf')}$, for observed
wavelengths on the blue side of the line: $\hat{\taures}(\lobs)$ = 
$-\ln$ ($\INPUT(\lobs)$ $(A')^{-1}$ $e^{\taudamp(\lobs, R_{S}',\nf' ) }
)$, which reduces to:

\vspace{+0\baselineskip}
\myputfigure{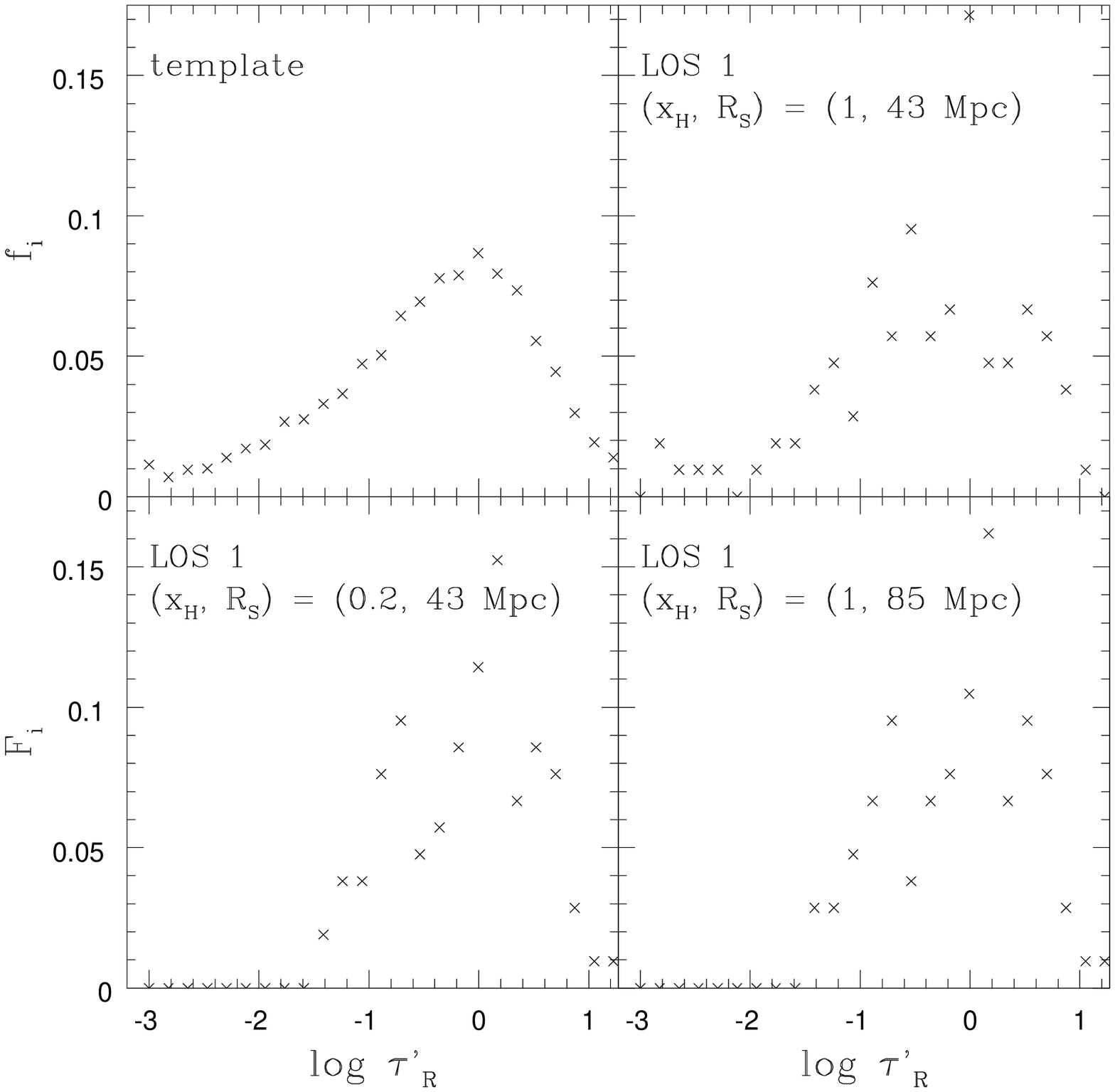}{3.3}{0.5}{.}{0.}  
\vspace{-1.1\baselineskip} \figcaption{Histogram of the template
distribution $\taures$ (\emph{top left}), and histograms of the
derived $\taures'$ distribution: $(\nf, R_S) = (1, 43 \rm ~Mpc)$
(correct values) (\emph{top right}), $(\nf, R_S) = (0.2, 43 \rm ~Mpc)$
(\emph{bottom left}), $(\nf, R_S) = (1, 85 \rm ~Mpc)$ (\emph{bottom
right}).  We test the hypothesis that the top right, bottom left and
bottom right histograms (among many others in parameter space) were
drawn from the distribution in the top left.  We find that the top
right panel is consistent with being drawn from the distribution in
the top left, and that the bottom panels are not.}
\vspace{+1\baselineskip}

\begin{equation}
\label{estimate_res_tau}
 \hat{\taures}(\lobs) = - \ln (e^{\taures(\lobs)} ~ e^{\taudamp' - \taudamp}) - \ln \left( \frac{A}{A'} \right)
\end{equation}

\noindent where $\taudamp' \equiv \taudamp(\lobs, R_{S}',\nf')$ and
$\taudamp \equiv$ $\taudamp(\lobs, R_{S}, \nf)$.  The first term on
the RHS of equation (\ref{estimate_res_tau}) represents our
misestimate of the damping wing contribution on the blue side, and the
second term represents our misestimate of the continuum inferred from
the red side.

Then, a final correction is applied to remove the contribution of the
assumed background flux corresponding to our guess of $\nf'$,
$J_{BG}'(\nu)$, to the hydrogen neutral fraction inside the \strom,
$x_H(z)$, defined by (\ref{neutralfraction}):

\begin{equation}
\label{final_estimate_res_tau}
\taures'(\lobs) = \hat{\taures}(\lobs) ~  \frac{\int_{\nu_H}^{\infty} (\frac{L_\nu}{4 \pi r^2} + 4 \pi J_{BG}') \frac{\sigma}{h\nu} d\nu}
{\int_{\nu_H}^{\infty} (\frac{L_\nu}{4 \pi r^2}) \frac{\sigma}{h\nu} d\nu}
\end{equation}

\noindent where $\taures'(\lobs)$ is our estimate for the resonance
optical depth at $\lobs$, for our guess of $(\nf', R_{S}')$, and
normalized to the resonance optical depth in a neutral universe
(i.e. $J_{BG} = 0$ in eq. [\ref{neutralfraction}]) for purposes of
comparison.

A histogram of $\taures'$ is constructed, and compared to the known
template histogram of $\taures$, extracted from the combined data from
92 smoothed mock spectra, created by embedding the source in a neutral
universe (see Figure 4).  In order to exclude spectral pixels that
would be too faint for an actual flux measurement, the inferred values
of $\taures' > 16$ were discarded from the analysis.  This choice is
motivated by the current limits available from the $z\sim 6$ SDSS
quasars.  The current best lower limit on the optical depth just
blueward of the Ly$\alpha$ line is actually worse ($\tau\sim 6$; White
et al. 2003), but since the higher Lyman series lines have lower
opacities, a better lower limit, $\tau\approx 22$, is available from
the Ly$\beta$ region of the spectrum (White et al. 2003). In
principle, the Ly$\beta$ (and Ly$\gamma$) spectral regions could be
added to our analysis.  A $\chisq$ test statistic is used to compare
the two templates:

\begin{equation}
\chisq = N_{points} \sum_{i=1}^{N_{bins}} \frac {(f_i - F_i)^2} {f_i}
\label{chisq_form}
\end{equation}

\noindent where $N_{points}$ is the number of $\taures'$ values
extracted from the spectrum, $f_i$ and $F_i$ are the fraction of
values expected and received in bin i, respectively.  The deviations
are approximately Gaussian, $\sigma_{i}^{2} = N_{points} f_i$.

This procedure is then repeated for many choices of $R_{S}'$ and
$\nf'$ (shown below), and the $\chisq$ test statistic in equation
(\ref{chisq_form}) is minimized.

\vspace{+1.1\baselineskip}
\myputfigure{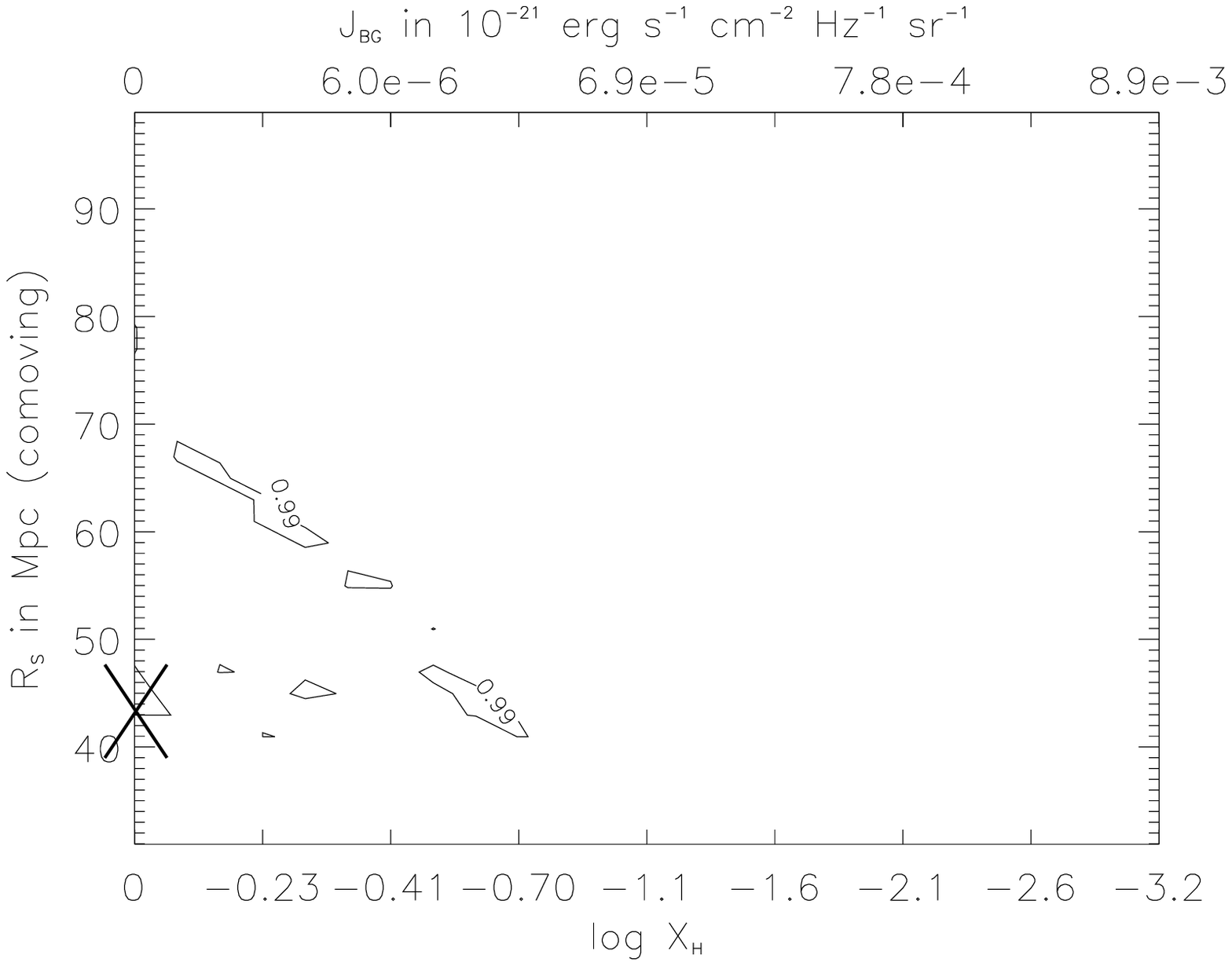}{3.3}{0.5}{.}{0.}  
\myputfigure{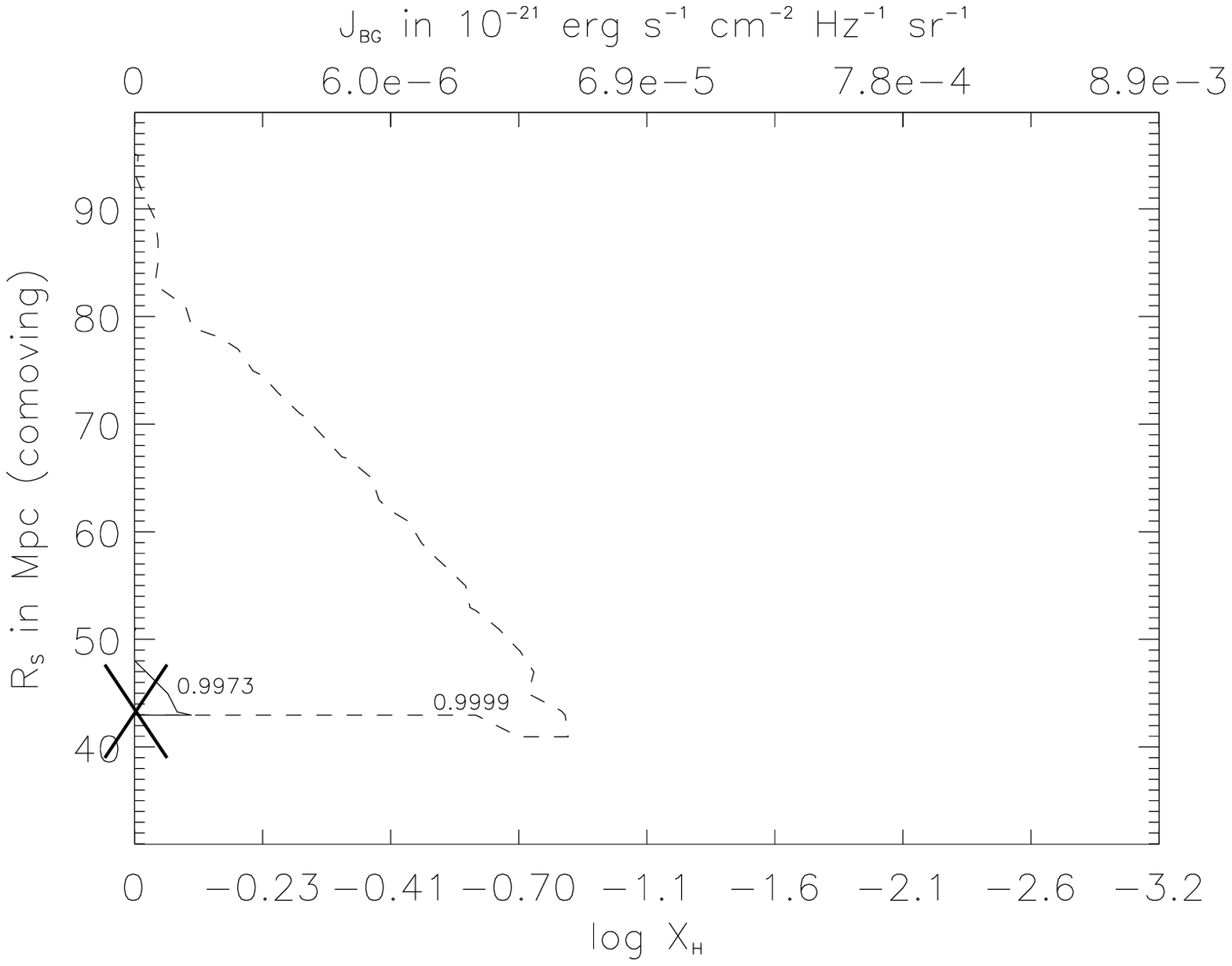}{3.3}{0.5}{.}{0.}
\vspace{-0.5\baselineskip} \figcaption{Probability contours for the
inferred neutral fraction and Str\"omgren sphere radius, generated by
the analysis in \S~3.1.  The background IGM is assumed to be neutral,
and the X marks the position of the true value of $R_S$ and
$\nf$. \emph{Top panel:} The inversion procedure is applied to a
single source.  The 90\% confidence contours (not shown) enclose only
the true ($\nf$, $R_S$) values, to within an error of (30\%, 5\%),
respectively 
 (only the single pixel containing the correct (fiducial)
values is enclosed within these contours).
  Confidence contours of 99\% are shown, and constrain
the IGM neutral fraction to $\nf \gtrsim 0.1$.  The irregular, island
shaped contours are the result of the unique features of a single LOS.
\emph{Bottom panel:} Probability contours generated by averaging the
same analysis over five random LOSs (averaging $\chisq$ values from
eq. [\ref{chisq_form}]) .  The irregular, 99\% confidence peaks from
the \emph{top panel} are smoothed out, and the contours are tighter.
The correct ($\nf$, $R_S$) values were chosen with 99\% confidence, to
within an error of (30\%, 5\%), respectively (contour not shown).  The
99.73\% and 99.99\% contours (\emph{solid} and \emph{dashed} lines,
respectively) are shown, ruling out a $\nf \lesssim 0.1$ universe at
better than 4$\sigma$ significance. }
\vspace{0.6cm}

The results of this procedure can be seen in Figures 5 and 6, and are
quite encouraging.  These figures show likelihood contours for
combinations of $(x_H', R_S')$ for two cases of the true parameters,
$(\nf, R_S)$ = (1, 43 Mpc) and $(\nf, R_S)$ = (0.008, 43 Mpc). The
statistical constraints shown in these figures are obtained from one
typical LOS to a single source.  Both cases had a minimum reduced
$\chisq \sim 1$ for 22 degrees of freedom ($d.o.f.$), corresponding to
two fitted parameters ($d.o.f. = (N_{bins} - 1) - 2$).  The $\nf
\times R_S$ probability grid in Figures 5 and 6 consists of $25 \times
35$ grid points, respectively.

In the case of a neutral universe (top panel of Fig. 5), the correct values of
$(\nf, R_S)$, to within an error of (30\%, 5\%), respectively, are
identified by our procedure at 90\% confidence.  We do not display
these contours, as they would be represented by a point in the figure.
Confidence contours of 99\% are shown, and constrain the IGM neutral
fraction to $\nf \gtrsim 0.1$.  The irregular, island contours in the
top panel of Fig. 5 are the result of the unique features of a single
LOS, and get smoothed out when averaged over several LOSs, as seen in
the bottom panel.  The bottom panel of Figure 5 shows the probability
contours, averaged over five different random LOSs.  The correct
($\nf$, $R_S$) values were chosen with 99\% confidence, to within an
error of (30\%, 5\%), respectively.  The 99.73\% and 99.99\% contours
(\emph{solid} and \emph{dashed} lines, respectively) are shown, ruling
out a $\nf \lesssim 0.1$ universe at better than 4$\sigma$, and an
$\nf \lesssim 0.7$ universe at 3$\sigma$.  As expected, confidence
contours get tighter as more LOSs are analyzed.

For the case of an ionized universe with $\nf \sim 0.008$ in Fig. 6,
the 1$\sigma$ uncertainty contours (not shown) enclose the correct
parameters, but the contours obtained from the single LOS are not as
tight as in the neutral universe case, as was predicted in the
introduction.  Nevertheless, a neutral universe was ruled out at the
99\% confidence level, as shown in Figure 6.

It is worthwhile to note that the iso-contours in Figures 5 and 6 go from bottom right to top left, indicating a degeneracy between small $\nf$,  small $R_S$, and large $\nf$, large $R_S$ solutions.  This is to be expected since the contribution of $\taudamp$ to the observed spectrum can be diminished both by moving the edge of the \stromspace further away from the source (increasing $R_S$), or by having a more highly ionized IGM (decreasing $\nf$).  Due to the limited spectral range used, the determination of the amplitude of the host's intrinsic emission, $A'$, is most affected by this degeneracy; however, since the shapes of the damping wings are different in these scenarios, the degeneracy is not exact and can be lifted by increasing the number of sources used in the analysis.

A useful by-product of this procedure is that it produces an estimate
of the amplitude of the intrinsic emission ($A' = A$ for the correct
choice of $(\nf, R_S)$).

\vspace{+1.1\baselineskip}
\myputfigure{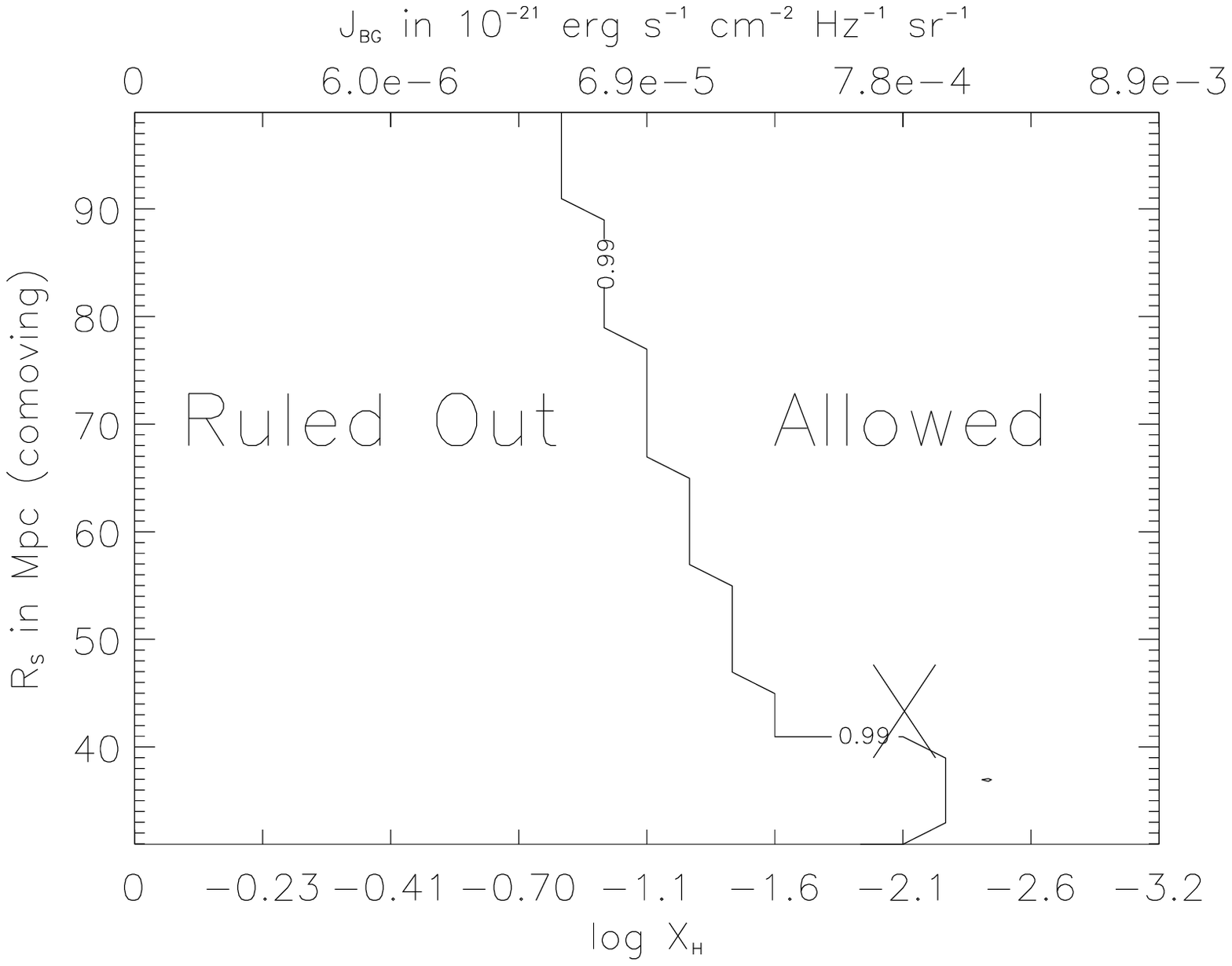}{3.3}{0.5}{.}{0.}  
\vspace{-0.5\baselineskip} \figcaption{ Probability contours for the
inferred hydrogen neutral fraction and Str\"omgren sphere radius,
generated by the analysis in \S~3.1.  The analysis is applied to a single source.  The background IGM is assumed to
have $\nf=0.008$, and the X marks the position of the true value of
$R_S$ and $\nf$.  Contours denote the 99\% confidence limits, and rule
out a neutral universe.  }
\vspace{+1\baselineskip}

An interesting issue of biasing can be raised here.  Since the quasar
sits in an overdense region, histograms of $\taures$ could have a
rather large scatter close to the line center due to the increased
spread in the density distributions and the peculiar motion of the gas
within the infall region around the halo (e.g., Barkana \& Loeb
2003).  This biasing effect was studied, and on average, we find that
the density field reaches within 15\% of the mean density $\sim$ 5
\AA\ blueward from the observed line center, for our redshift
$\zsource=6$ quasar.  However, the distribution of densities at a
fixed distance has a high-end tail, comprised of a few LOSs.  Although
we can model this bias for our quasar, one would need the density
profiles surrounding many host sources to fully explore these
statistics for a larger sample of quasars.  This is feasible in principle, in large future
simulations. However, for the purposes of generality, in the present
paper the blue side of the spectra was cut-off 5 \AA\ blueward of the
line center and omitted from our analysis.  Environment biasing will be discussed in detail in \S~\ref{sec:bias}.

Another uncertainty can arise if the redshift of the \lya line
center is not well determined.  An offset in the line center of $\pm
1000 ~ \rm km ~ s^{-1}$ (typical of quasar jets), results in an
effective $R_S$ offset of only $\pm$ 1.3 comoving Mpc.  A larger
systematic offset of $\Delta z \sim 0.01-0.05$ in the line center,
corresponding to $\Delta R_S \sim 4-20$ Mpc, can be present if an
associated metal line with a well--determined redshift has not been
detected.  This can represent a non--negligible misestimate in the
radius of the ionized region (especially for faint sources), and
underscores the importance of accurate redshift--determinations from
metal--line detections.  Nevertheless, we note that even these
relatively large $R_S$ uncertainties do not have a major impact in
distinguishing among the $\nf$ values of interest, which are separated
by two orders of magnitude.  This can be seen explicitly by noting
that a vertical change of order $\Delta R_S \sim 4-20$ Mpc along the probability iso-contours, seen in
Figures 5 and 6, corresponds to less than an order of magnitude change in the
value of the neutral fraction.  This is the case even for fainter
sources, as discussed below.

\subsection{Simulating Error in the Emission Template}

The assumption of knowing the shape of the quasar's intrinsic emission
is not as outlandish as it might seem.  \citet{vb01} created a
composite quasar spectrum from $\sim 150$ SDSS spectra of quasars
around $z \sim 3$.  Even though there were variations in amplitude,
the normalized, mean template was defined around the \lya line to
better than 5\%.  Although spectrum-to-spectrum variations are
considerable in the core of \lya and blueward, most of these
variations are accounted for by the Baldwin effect (the decrease of
emission line width with increasing luminosity) and the absorption by
the \lya forest, both of which can be statistically removed.  Hence,
for this paper, we assumed spectrum-to-spectrum variations of
$\lesssim 20\%$; a typical value of the r.m.s. scatter in the
continuum level immediately redward of \lya \citep{vb01}.

Fan et al. (2003) demonstrate that the high-redshift quasars follow
\lya emission line and continuum spectral shapes that are remarkably
similar to that of their low--redshift counterparts.  More precisely,
the high redshift sample size is small, but the spectra of the present
handful of $z \sim 6$ quasars are consistent with being drawn from the
Vanden Berk et al. distribution of spectral shapes (Fan 2003, private
communication).

Here we address the effect of the uncertainties of the assumed
spectral template in two ways: we assume (1) an unknown overall
``tilt'' in the spectrum, or (2) an unknown pixel--to--pixel,
uncorrelated, normally distributed scatter around a well--determined
mean spectrum.  These uncertainties add extra parameters in our
analysis, and make the constraints on the neutral fraction less tight.
However, as we will argue below, tight statistical constraints on the
neutral fraction can still be obtained with a sufficiently large
sample of high redshift sources.

\subsubsection{Tilt in the Emission Template}

The uncertainty in the emission template was first modeled with a
pivoting procedure, chosen to be hinged at the wavelength used for the
amplitude estimation in eq. (\ref{probed_A}), $\lambda_{piv}$ =
$\lambda_{red}$ = 8514 \AA.  This mimics an incorrectly chosen
power--law for the quasar's continuum emission (and also a tilt in the
\lya emission line profile).  Such a tilt would also characterize an
uncertainty in the power--law index of GRB afterglow spectra (see
\S~5.2).

Thus, the observed flux is assumed to be of the form: $\INPUT(\lobs)$ = $A$~$
(\frac{\lobs}{\lambda_{piv}})^{\alpha}$~$e^{-\taures(\lobs) -
\taudamp(\lobs, R_{S},\nf)}$.  Furthermore, equation
(\ref{estimate_res_tau}) is modified to:

\begin{equation}
\label{estimate_res_tau_tilt}
\hat{\taures}(\lobs) = - \ln (e^{\taures(\lobs)} ~ e^{\taudamp' -
  \taudamp}) - \ln \left( \frac{A}{A'}
\left(\frac{\lobs}{\lambda_{piv}} \right)^{\alpha - \alpha'} \right)
\end{equation}

\noindent where $\alpha'$ is now a guess for the spectral slope $\alpha$.

The shape of this power--law tilt is similar, but not mathematically
degenerate with the shape the damping wing. It therefore causes
constraints to degrade. Here we characterize this degradation using:

\begin{equation}
\langle N_{tilt} \rangle \approx \frac{1}{\sigma_{tilt}^2}
\label{tilt_num}
\end{equation}

\noindent where $\langle N_{tilt} \rangle$ is an estimate of the
number of data points (i.e. spectral resolution bins) required to
distinguish between the two spectrum shapes, and $\sigma_{tilt} \equiv
\Delta \tau_{D-t} / \Delta \taures$, with $\Delta \tau_{D-t} \equiv
\langle | \ln (\frac{A}{A'}e^{-\taudamp + \taudamp'}) - \ln
(\frac{\lobs}{\lambda_{piv}})^{\alpha - \alpha'} | \rangle$ being a measurement
of the typical spread of the estimated $\taures'$ obtained from
differences in the shape of the power--law and the shape of the damping
wing flux decrement, and $\Delta \taures \approx 0.1$ is the width of
the template $\taures$ histogram.

We find that the tilt in the emission template obtained with a
misestimate of $\alpha$ is not degenerate with the shape of the
damping wing for our quasar.  The largest value of $\langle N_{tilt}
\rangle$, obtained from our parameter space occurred for the
degeneracy between the fiducial values of ($\nf=0.008$, $R_S=43$ Mpc)
and the parameter choice of ($\nf'=1$, $R_S'=99$ Mpc) with a tilt
misestimate of $\alpha - \alpha' \approx -3$.  For these values,
$\langle N_{tilt} \rangle \sim 10^3$, corresponding to, e.g., 10
similar quasar spectra with 100 usable independent spectral resolution
elements.

\subsubsection{Pixel-by-Pixel Errors}

The uncertainty in the emission template was next modeled assuming
uncorrelated, Gaussian distributed errors.  Such pixel--by--pixel
uncertainties could also represent the noise associated with the flux
detection in each bin.  Hence, the value of the emission template,
$T(\lobs)$, in the input flux in equation~(\ref{input_flux}), now
becomes a Gaussian distributed random variable with a mean value for
each wavelength given by $\langle T(\lobs) \rangle$.  The inversion
equation to replace equation~(\ref{estimate_res_tau}) then becomes:

\begin{eqnarray}
\label{gaus_res_tau}
\hat{\taures}(\lobs) = &-& \ln (e^{\taures(\lobs)} ~ e^{\taudamp' - \taudamp}) - \ln \left( \frac{A}{A'} \right)\\
\nonumber &-& \ln \left( \frac{T(\lobs)}{\langle T(\lobs) \rangle} \right)
\end{eqnarray}

\noindent In the case of small deviations from the template value,
$\frac{T(\lobs)}{\langle T(\lobs) \rangle}$ $\sim$ 1,
$\hat{\taures}(\lobs)$ is approximately Gaussian distributed around [-
$\ln (e^{\taures(\lobs)}$ $e^{\taudamp' - \taudamp}) - \ln \left(
\frac{A}{A'} \right)]$.  The exact shape of the probability
distribution of $\taures$ resulting from the Gaussian uncertainty in
$T(\lobs)$ was calculated and convolved with the template $\taures$
distribution (top left panel in Fig. 4).  This new, normalized,
template histogram was used as the new probability density function,
$f_i$, in the $\chisq$ statistic shown in (\ref{chisq_form}).

To test the robustness of our results to these pixel--by--pixel errors,
a random LOS was drawn from our pool of LOSs, a Gaussian distributed
emission template, $T(\lobs)$, was generated to create the mock
spectrum, and the inversion procedure outlined in \S~4.1 was
performed, using equation (\ref{gaus_res_tau}) to generate
the $\taures'$ histograms.  This procedure was repeated, updating the
$\chisq$ values in the parameter space with each newly processed LOS
(each new LOS representing a different source in a hypothetical
sample) until we were able to distinguish a neutral universe from a
$\nf < 0.008$ universe with 99\% confidence.

The results for the number of sources needed for the 99\% confidence
constraints are summarized in Table \ref{tbl1}.  The emission uncertainties refer to the standard deviation of the pixel--by--pixel Gaussian distributed errors in the emission template.  With a 20\%
uncertainty in the intrinsic emission template, an average of,
$\langle N_{LOS} \rangle \sim 34$ LOSs were required to rule out a
neutral universe when the true value of the hydrogen neutral fraction
was $\nf = 0.008$.  In the $\nf = 1$ regime, only $\sim$ 3 spectra on
average were required to rule out a $\nf < 0.008$ universe with 99\%
confidence.  This should be expected, since as discussed previously,
the neutral IGM leaves a heavier footprint on the quasar spectra for a
reasonably sized \strom.

Aside from the implicitly assumed calibration and elimination of the
Baldwin effect, the adopted 20\% uncertainty does not make use of any
information (such as correlations with other observables) from the
observed spectrum being processed to further improve constraints.
Preliminary results from a principal component analysis (PCA) of the
spectrum--to--spectrum variations suggest that one might be able to
characterize the variance in the emission template with only three
eigenvalues (Vanden Berk 2003, private communication).  It is likely
that with such characterization of the variance and the usage of
information redward of Ly$\alpha$, the template uncertainty can be
reduced, thus reducing $\langle N_{LOS} \rangle$.  With a 10\%
pixel--by--pixel uncertainty, an average of 25 LOSs were needed to rule
out a neutral universe for $\nf = 0.008$, and 3 to rule out an $\nf <
0.008$ universe for $\nf = 1$.  A 5\% uncertainty brings the average
number of required spectra down to 14 and 2, for $\nf = 0.008$ and
$\nf = 1$, respectively.  These results, as well as the
r.m.s. deviations, are summarized in Table \ref{tbl1}.

\begin{table}[ht]
\caption{Average number of bright quasars required to distinguish
between $\nf = 1$ and $\nf < 0.008$ with 99\% confidence.}
\vspace{-0.5cm}
\label{tbl1}
\begin{center}
\begin{tabular}{ccc}
\tablewidth{3in}\\
\hline
\hline
Emission & $\langle N_{LOS} \rangle$ with & $\langle N_{LOS} \rangle$ with\\
Uncertainty & $\nf = 1$ & $\nf = 0.008$\\
\hline
20\% & 3.42 $\pm$ 1.95 & 33.6 $\pm$ 20.8\\
10\% & 3.39 $\pm$ 2.15 & 24.6 $\pm$ 20.0\\
5\% & 2.45 $\pm$ 1.13 & 16.8 $\pm$ 14.2\\
\hline
\end{tabular}\\
\end{center}
\end{table}

\section{Merits of Different Source Types}
\label{sec:issues}

Although the preceding analysis was done with simulated quasar spectra
similar to the bright SDSS high--redshift quasars, it can be repeated
on other high-redshift sources, namely quasars with lower
luminosities, high--redshift galaxies and GRB afterglows.  The
properties of the sources scale predictably with their luminosities,
with the two most pertinent to this analysis being the size of the
source's Str\"omgren sphere and the shape of the source's intrinsic
emission spectrum.

\subsection{Str\"omgren Sphere Size}
The size of the source's Str\"omgren sphere, $R_S$, approximately
scales as the source's $\rm (luminosity)^{1/3}$.  A fainter quasar, of
similar age, might therefore have fewer measurable spectral points
inside the Str\"omgren sphere, with which to create the $\taures'$
histogram, which weakens constraints.  On the other hand, since the
Str\"omgen sphere is smaller, the damping wing has a stronger effect
on the spectrum, which strengthens constraints.

To estimate the overall effect on the determination of the neutral
fraction, we repeated the pixel--by--pixel analysis outlined in \S~4.2.2
on a 100 times fainter quasar ($L_\nu$ = 2.34 $\times$ $10^{29}$ $(\nu/\nu_H)^{-1.8}$ 
$[(1+z)/(1+\zsource)]^{-0.8}$) with a correspondingly smaller
Str\"omgren sphere ($R_S$ $\approx$ 9.3 Mpc).  From simulated spectra of such mock
quasars in a neutral universe, we find that a $\nf < 0.008$ universe
is ruled out at 99\% confidence with $\sim$ 3 LOSs.  This result is
therefore comparable to that available from the brighter quasars
(shown in Table 1).  This can be understood by realizing that even though fewer spectral points are available for the analysis of a fainter source (since the \strom\ is smaller), these points are those close to the edge of the \strom\ where the damping wing has a sharper slope, and are given more statistical weight due to the lack of points far away from the edge of the \strom, where the damping wing is flatter.
On the other hand, we find that approximately
$10^2$ - $10^3$ faint quasar LOSs are needed on average, for the emission template
uncertainties of 5\%-20\%, in order to break the degeneracy between
the damping wing of a small $\nf$, small $R_S$ and a large $\nf$,
large $R_S$ parameter choice.  Since the probability iso-contours go
from bottom right to top left in the parameter space of Figure 5,
limiting parameter space to $R_S < 100$ Mpc (as we had implicitly done
above) allowed for a stronger degeneracy in an ionized universe in the
case of a small $R_S$ than in the case of the larger $R_S$ quasar used
throughout the preceding analysis in this paper.  If instead we limit
parameter space to $R_S < 30$ Mpc, the fainter sources are able to
distinguish between a neutral universe and a $\nf < 0.008$ universe,
using again a comparable number (i.e., tens) of sources to the
brighter quasars shown in Table 1.  Whether limiting the size of the
Str\"omgren sphere to 30 Mpc is a reasonable prior assumption depends
on the actual inferred neutral fraction. The size of the Str\"omgren
sphere scales as $x_H^{-1/3}$, and for sources that are $\sim 100$
times fainter than the $z\sim 6$ SDSS quasars, they would require
$\gsim 10^7$ years to reach a 30 Mpc size even if embedded in an IGM
with neutral fractions as low as $x_H=0.008$.

An implicit drawback of fainter sources is that they have smaller S/N
than bright sources, thereby reducing the effective detection
threshold.  \citet{becker01} and \citet{white03} state the 1$\sigma$
lower limits to the \lya optical depth inferred from the \lya trough
to be $\tau_{lim} \sim$ 5 or 6, for the Keck ESI spectra of z $\sim$ 6
quasars.  As stated previously, the analysis in this paper ignored
inferred values of $\taures' > 16$.  Because of the sharp drop in the
high--end tail of the $\taures$ histograms (see Fig. 4), the inversion
procedure presented in \S~\ref{sec:invert_spect} is not sensitive to the
exact value of the detection limit.  In our template histogram, only
5\% of the values were within the range $6 < \taures < 16$,
approximately corresponding to data points used in the analysis, which
are undetectable with $\tau_{lim} \sim 6$.\footnote{There is a small
bias associated with the fact that the detection limit, $\tau_{lim}$,
is a limit on the total optical depth, $\tau \equiv \taures +
\taudamp$, and not just $\taures$.} However, a 100 times fainter
quasar would have its optical depth detection threshold reduced by
ln(100)=4.6 to obtain equivalent S/N for the same integration time.
For our mock spectra, this would eliminate an additional 20\% of the
$\taures$ values, in the range of $1.4 < \taures < 6$.  Hence, fainter
sources have a smaller range of usable $\taures$ values than brighter
sources.  Nevertheless, higher S/N observations would decrease this
effect.

Sources that are too faint to have many detectable spectral points on
the blue side of the line, such as faint galaxies and GRB afterglows,
would not be useful in this $\taures$ histogram analysis, but would
still have a strong damping wing imprint on their spectrum redward of
Ly$\alpha$.  As mentioned previously, redward of Ly$\alpha$, $\taures$
is negligible, and the small-scale power analysis presented in \S~4.1
is irrelevant.  If the sources are intrinsically faint enough, the
slope of the damping wing redward of \lya is sufficient to distinguish
between a small $R_S$, small $\nf$ case and a large $R_S$, large $\nf$
scenario with only a few sources.  As emphasized recently by
\citet{lh03} and Barkana \& Loeb (2004), GRB afterglows with a clean
power--law intrinsic spectrum would be especially well suited to such
an analysis, provided they can be identified in future datasets
(e.g. of Swift).

As an example, the usefulness of the red side of the \lya line in
determining $\nf$ was also investigated for our simulated faint quasar
with $R_S \approx 9.3$ and $\nf \approx 0.008$, assuming
pixel--to--pixel Gaussian scatter in the emission template.  In
principle, sufficiently far on the red side of the Lyman $\alpha$
line, where resonant absorption does not contribute, this analysis
could be performed without a numerical simulation.  However, due to the
cosmic infall, some of the pixels immediately on the red side of the
\lya line center correspond to foreground gas between us and the quasar
(the resonant absorption extends, in our case, up to 7-8 Angstroms to
the red, see Figure 8).  For this reason, we did use the simulation in
the analysis. Alternatively, to avoid uncertainties due to modeling
the cosmological infall, one could omit the spectral regions that are
blueward of the infall regime. Here we utilized the flux in the
observed wavelength range of 8514 \AA\ -- 8544 \AA\ and performed a
simple comparison. A more robust analysis of the red side would
involve $\chisq$ minimization in the 3-dimensional parameter space of
($\nf$, $R_S$, $A$), where $A$ is the amplitude of the intrinsic
emission.  However, for purposes of comparison to the $\taures$
histogram analysis presented in
\S~\ref{sec:invert_spect}, it was sufficient to obtain an estimate of
$A$ using an assumed region in the observed spectrum with an
accurately known template shape, and calculating $A'$ as outlined in
\S~3.1 and equation (\ref{probed_A}).  For a choice of $\nf'$ and
$R_S'$, we therefore obtain an estimate of the flux decrement due to
resonance absorption, $d_R(\lobs) = \INPUT(\lobs) ~ \frac
{e^{\taudamp'}} {A' \langle T(\lobs) \rangle}$, which expands to:

\begin{equation}
d_R(\lobs) = e^{-\taures(\lobs)} ~ \frac{A}{A'}  \frac{T(\lobs)}{\langle T(\lobs) \rangle}  e^{\taudamp' - \taudamp}
\label{d_R_equation}
\end{equation}

\noindent For the correct parameter choices, $d_R(\lobs)$ $\approx$ 
$e^{-\taures(\lobs)}$ $\approx$ 1, since $\taures \rightarrow 0$ on the
red side of the line and redward of the infall regime.  The mean flux
decrement, $\langle d_R(\lobs) \rangle$, averaged over several
randomly chosen LOSs, was compared to the average decrement obtained
from the simulated spectra at each wavelength, $\langle
e^{-\taures(\lobs)} \rangle$, with the $\chisq_R$ statistic:

\begin{equation}
\chisq_R = \sum_{l_{\rm min}}^{l_{\rm max}} \frac {(\langle e^{-\taures(\lobs)} \rangle - \langle d_R(\lobs) \rangle)^2} {\sigma_m^2}
\label{chisqres_form}
\end{equation}

\noindent where the summation limits were chosen to be $l_{\rm min} =
8514$ \AA\, $l_{\rm max}$ = 8544 \AA\ and $\sigma_m$ is the standard
deviation of the intrinsic emission template divided by the square
root of the number of LOSs used in determining the mean decrement,
$\langle d_R(\lobs) \rangle$.  LOSs were chosen at random, and the
mean decrement and test statistic in equation (\ref{chisqres_form})
were updated until a neutral universe could be ruled out at 99\%
confidence.

Results from this procedure are quite comparable to the $\taures$
histogram analysis on the blue side of the Ly$\alpha$ line for our
faint quasar with $R_S \approx 9.3$ and $\nf \approx 0.008$.  On
average, we find that $10^2$ -- $10^3$ LOSs were needed to rule out a
neutral universe at 99\% confidence for template uncertainties of
5\%-10\%, just as in the $\taures$ histogram analysis.  Although the
red side is cleaner (negligible resonant absorption), it is further
away from the edge of the \strom, so the damping wing shape is flatter
and is therefore harder to detect.  However, our results suggests that
if a larger wavelength range is available for the analysis, or if the
source has a smaller \strom $~(\sim$ a few Mpc), analyzing the red
side of the \lya line could prove more efficient than the blue side in
determining $\nf$.\footnote{There is an additional complication that
an intervening damped Lyman alpha (DLA) system along the line of sight
causes absorption whose shape is partially degenerate with that due to
the damping wing from a smooth IGM (Miralda-Escud\'e 1998; Barkana \&
Loeb 2004).  However, the degeneracy is not exact, and lines of sights 
with DLAs should be rare.}

Furthermore, very faint sources have other tracers of $\nf$.  For
example, \citet{zh02} showed that the shift in the peak of the
observed \lya emission line (relative to other emission lines), or its
measured asymmetry, as a function of the source luminosity, can be
used as a probe of the IGM hydrogen neutral fraction.

\subsection{Intrinsic Emission}

Our knowledge of the source's intrinsic emission varies considerably with
source type.  Again, it is only the imperfect knowledge of the shape
of the emission spectrum that complicates the analysis presented here
(i.e. knowledge of the overall amplitude of the spectrum is not
needed).  On average, quasars seem to have a moderately well
constrained emission template around Ly$\alpha$, with an uncertainty
of $\sim 20\%$, once the spectrum is normalized and systematics such
as the Baldwin effect are accounted for.  It is possible that this
uncertainty could be further decreased with a principal component
analysis (PCA), which has yielded well--characterized
spectrum--to--spectrum dispersions on the red side of the Ly$\alpha$
line (Vanden Berk 2003, private communication).  In comparison, the
spectral shape of emission from galaxies is more poorly defined around
\lya \citep{steidel01}. One would therefore need larger samples of
Ly$\alpha$ emitting galaxies at both lower redshifts and higher
redshifts for a better empirical calibration of their spectral shape
and its dispersion.  Such samples may soon be available from extensive
Ly$\alpha$ searches using HST (Rhoads et al. 2003) and Subaru
(Taniguchi 2003).  GRB afterglows have a smooth power--law spectrum
shape.  There appears to be a scatter in the power--law index of
roughly $\sim$ 10\% in photometric data on low--redshift bursts. The
sample of higher--redshift bursts, whose spectral shape can be studied
in more detail around the Ly$\alpha$ line, is still too small to
quantify spectrum-to-spectrum dispersions, but they appear to closely
follow power--laws \citep{miretal03}. Note that a pure power--law
uncertainty is not degenerate with the shape of the damping wing (see
\S~4.2.1).  The major issue with GRB afterglows will be whether the
high--redshift afterglows can be identified among the lower--redshift
events (see, e.g., Lamb \& Reichart 2000 and Barkana \& Loeb 2004 for
recent discussions).

\subsection{Environment Bias}
\label{sec:bias}

The source's host environment has a large effect on the observed
spectrum near the line center, as emphasized recently by Barkana \&
Loeb (2004). To accurately model this effect, and to test
semi--analytical models \citep{barkana04} hydrodynamical simulations
are needed to generate the local density and velocity fields.  In
Figures 7 and 8, we show the effect of this biasing for our mock
quasar.  Figure 7 shows the histograms of the spherically averaged
(within our set of 92 lines) radial component of the peculiar velocity
fields constructed from 2 \AA\ bins in $\lobs$, at increasing
distances away from the quasar.  Most of the radial velocities closest
to the host pixel exhibit strong infall, with the PDF peaking at
peculiar radial velocities of $v \approx$ -100 km/s.  However, there
is a significant, high-velocity tail in the distribution,
corresponding to several LOSs that exhibit strong outflows (100-500
km/s) close to the host pixel.  As the distance from the central
overdensity increases, the radial velocity PDFs become narrower and
more centered around 0 km/s, as is expected for a randomly chosen
point in space.  Strong outflow features disappear $\sim$ 5 \AA\ away
from the line center.  After about 10 \AA\ away from the line center,
the velocity histogram becomes quite symmetric around 0 km/s.

These proximity effects have a strong impact on the resonance optical
depth, $\taures$, as shown in Figure 8.  The dotted curve shows
$\taures$ calculated assuming the mean density, $n_{H}(z) =
n_{H,0}~(1+z)^3$, and not including velocity information from the
simulation.  The dashed curve includes the density field from the
simulation, but not the velocity field.  Finally, the solid curve
includes both the density and velocity information from the simulation
box.  The values of $\taures$ shown in this figure were averaged over
all LOSs: $ -\ln \langle e^{-\taures}\rangle$.  The figure clearly
shows that the peculiar velocities around the density peak smooth out
the spectrum, and create a more gradual decline in the optical depth
redward of Ly$\alpha$.  Overdensities close to the line center become
regions of relatively low optical depth, since the gas Doppler shifts
out of resonance.  Further away on the blue side from the host
overdensity, the scatter in the density and velocity fields among
different LOSs decreases, and the velocity histogram becomes centered
around 0 km/s (as mentioned above), so the difference between the
averaged $\taures$ curves becomes statistically negligible.

These figures merely emphasize that density biasing due to the host
environment is important, and it requires a larger simulation with a
statistical sample of density peaks to quantify. We plan to carry out
such an analysis in detail in a future paper.  In particular, the
analysis presented in this paper should be performed for a larger
number of density peaks in other simulation boxes for the density bias
to be statistically analyzed.  In actual data analysis, it should help
that the host environment should scale with the halo size, and
therefore with the source's luminosity.

\vspace{+1\baselineskip}
\myputfigure{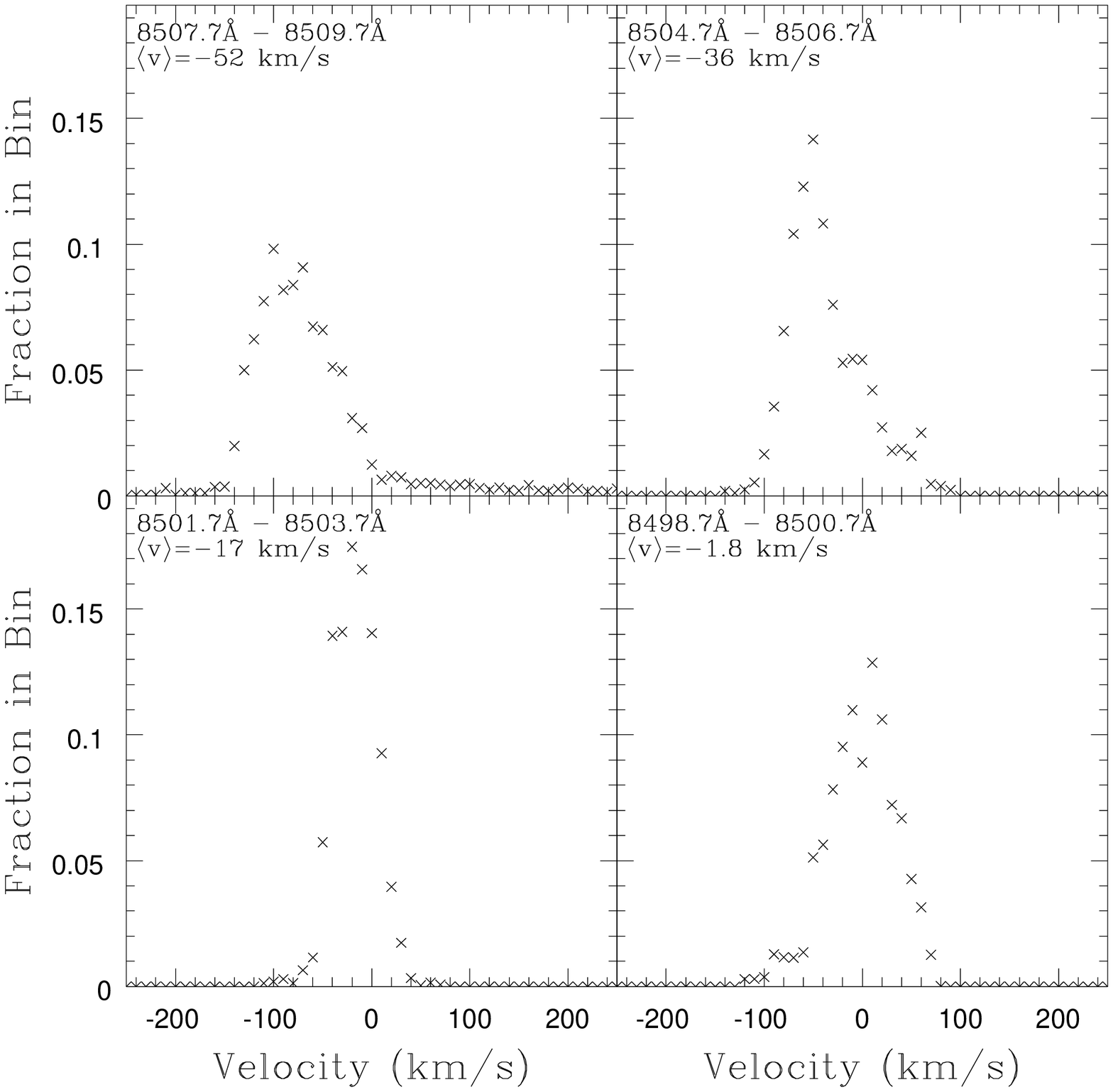}{3.3}{0.5}{.}{0.}  
\vspace{-1\baselineskip} \figcaption{Spherically averaged radial
velocity histograms at several distances, corresponding to redshifts
of $z$ = $\lobs$/$\lambda_\alpha$ - 1, away from the host pixel at $\zsource=6$.
The histograms are constructed from different $\lobs$ ranges:
8507.69\AA\ - 8509.69\AA\ (\emph{top left}), 8504.69\AA\ - 8506.69\AA\
(\emph{top right}), 8501.69\AA\ - 8503.69\AA\ (\emph{bottom left}),
and 8498.69\AA\ - 8500.69\AA\ (\emph{bottom right}).  The fraction of
infalling gas decreases with increasing distance from the central
overdensity, until at about 10 \AA\ away from the line center, at
which distance the histogram becomes nearly symmetric around 0 km/s.
Close to the central overdensity (\emph{top left}), most of the gas
has large negative radial velocities, but there is a significant,
high-velocity tail in the distribution.  }
\vspace{+1\baselineskip}

\vspace{+1\baselineskip}
\myputfigure{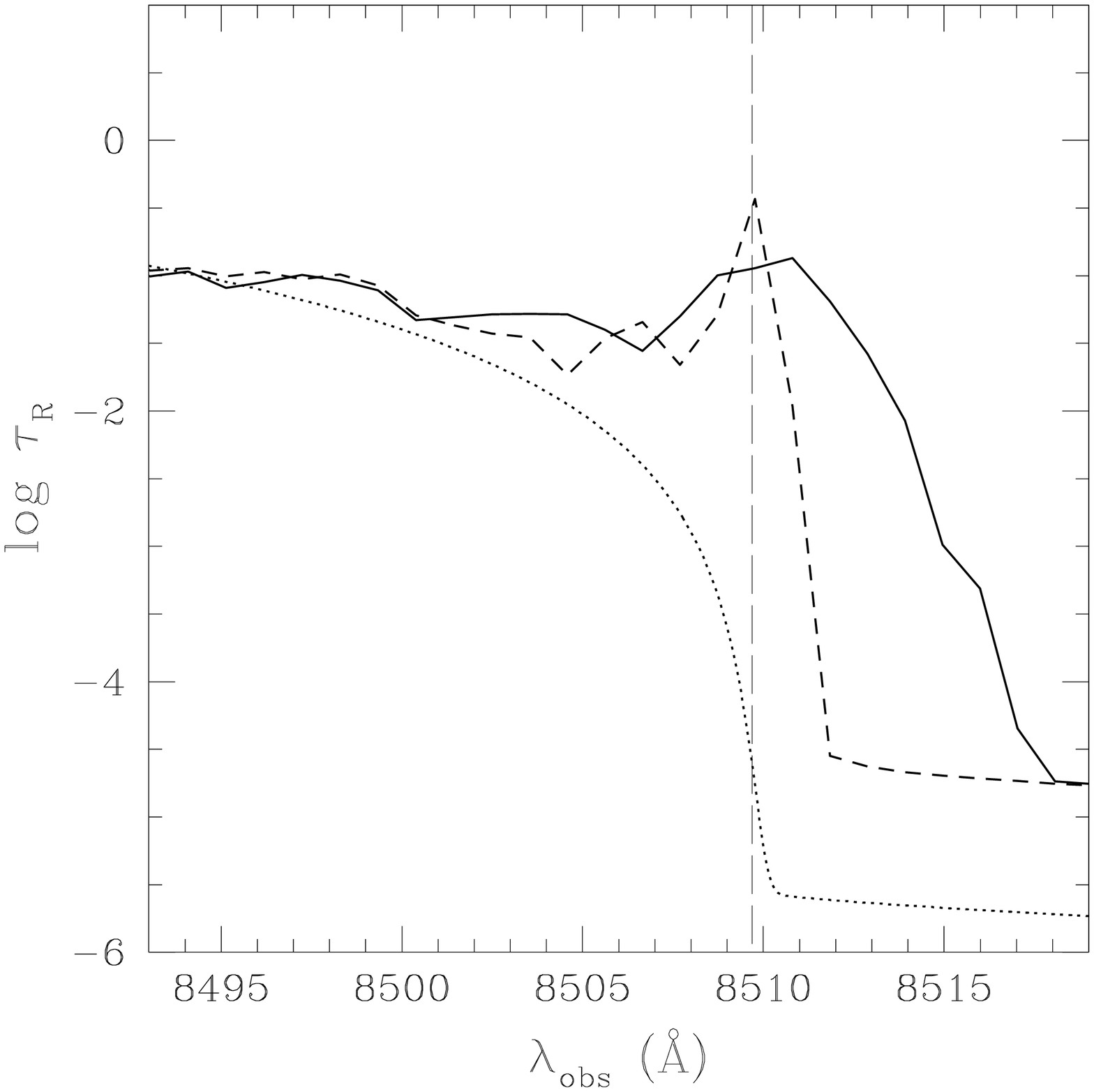}{3.3}{0.5}{.}{0.}  
\vspace{-1\baselineskip}
\figcaption{
Values of $\taures$ averaged over all LOSs: $-\ln \langle
e^{-\taures}\rangle$.  The \emph{dotted curve} shows $\taures$
calculated assuming the mean IGM density, $n_{H}(z) = n_{H,0}~(1+z)^3$,
and not including velocity information.  The \emph{dashed curve}
includes the density field from the simulation, but not the velocity
field.  The \emph{solid curve} includes both the density and velocity
information from the simulation box.  The \emph{long dashed vertical
line} denotes the \lya line center at 8509.69 \AA.  The solid curve
can be approximately reproduced by convolving the dashed curve with
the velocity PDFs at each $\lobs$.  Including velocity information
smooths out the spectrum, and creates a more gradual decline in the
optical depth redward of Ly$\alpha$.  Further away on the blue side
from the host overdensity, the difference between the averaged
$\taures$ curves becomes statistically negligible.}
\vspace{+1\baselineskip}

On the other hand, the biased region does not encompass the
majority of the spectrum, as mentioned in \S~\ref{sec:invert_spect}.
For our mock quasar, the biased region extends only $\sim$ 5 \AA\ away
from the line center.  However, since our halo is smaller than that
expected to host typical $z\sim 6$ quasars, the biased region could be
larger for a more realistic, more massive halo. Using a
semi--analytical model, Barkana \& Loeb (2004) recently derived the
gas density and velocity distributions around the relevant halos.
They find (see their Figure 1) that for halo masses typical of those
expected to host the bright SDSS quasars, the biased region should
extend about $\sim$ 1 Mpc (proper).  This translates to a wavelength
range of $\sim$ 20 \AA\, and still influences less than 1/6 of the
region used in our analysis.  Our initial results therefore suggest
that one can extract $\taures'$ histograms from spectra even without
modeling the density bias, by ignoring this part of the spectrum
around the line center, as was done in this paper.

It is interesting to ask which part of the optical depth distribution
carries the most statistical power - the low, or the high--optical
depth pixels, since the low--optical depth pixels are preferentially
effected by biasing effects.  We addressed this issue by computing the
statistical power of fractions of our $\taures$ histograms. We found
that similar constraints are obtained by using only pixels with
$\taures <$ 0.1 and by using only pixels with $\taures >$ 0.1 (see
Fig. 4).  Approximately 1/3 of the underdense ($\taures <$ 0.1) pixels
used in our analysis are expected to lie within the biased region
discussed above. It would clearly be beneficial to have accurate
statistics of the biasing of host halos, but this is impractical given
current computational constraints.  The simplest way to deal with the
issue is to remove the biased region from the analysis (as was attempted in this paper).  
Since the majority of pixels are not
affected, as discussed above, we expect our results to be fairly
insensitive to biasing.

As a final test of local bias, we created a new template distribution
of $\taures$ values by using LOSs generated from a new and independent
simulation box.  The densest region in the new box, from which LOSs
originated, corresponds to a dark matter halo of mass $M_{halo}
\approx 1\times 10^{10} M_{\sun}$, about 1/2 the mass of the densest
halo in the simulation box used throughout the rest of this paper.
The two template distributions can be seen in Fig. 9, with {\it
squares} representing the new histogram, and {\it crosses} showing the
old histogram (i.e. the top left panel of Figure 4).  The two
histograms are fairly similar (their difference is much smaller than
those between the models being compared to one another in Figure 4).
Our old template is somewhat wider, as would be expected from the
larger density contrast we find in that simulation box.  The density
field around the new halo reaches within 15\% of the mean density
$\sim$ 5 \AA\ blueward from the line center, just as in the former
halo.  

\vspace{+1\baselineskip}
\myputfigure{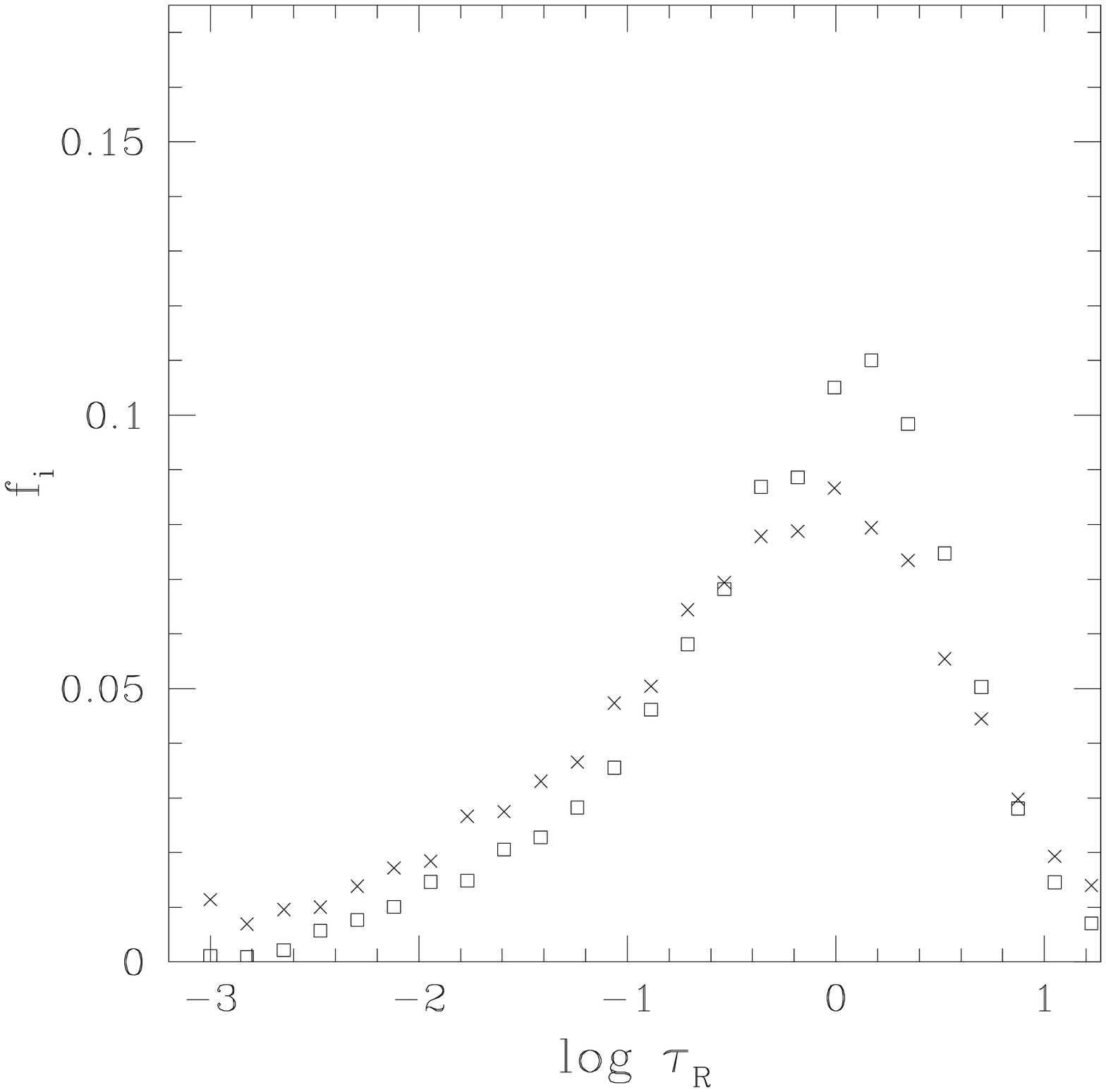}{3.3}{0.5}{.}{0.}  
\vspace{-1\baselineskip} \figcaption{Histogram of the template distribution of $\taures$ from LOSs originating at a dark matter halo of mass $M_{halo} \approx 2\times 10^{10} M_{\sun}$ ({\it crosses}), and originating at a dark matter halo of mass $M_{halo} \approx 1\times 10^{10} M_{\sun}$ ({\it squares}).  The former distribution was used throughout this paper.  The later distribution was created from a different simulation box, in order to investigate uncertainties in the template.  The two distributions are similar, with the histogram corresponding to a larger mass host halo being somewhat wider due to the larger density contrast in the simulation box.}
\vspace{+1\baselineskip}

\section{Adding Additional Constraints: Limits on $\mbox{R}_S$}
\label{sec:r_limits}

As discussed above, the dominant factor in this analysis is the
degeneracy between small $\nf$, small $R_S$, and large $\nf$, large
$R_S$ parameter choices.  Hence, it would be quite useful to be able
to independently restrict allowed values of $R_S$.  This is difficult
since the mean resonant optical depth grows roughly as the square of
the distance away from the source, so resonance absorption inside the
\strom\ can be sufficient to block out \lya flux alone, especially
considering uncertainties in the source's ionizing luminosity.

However, a particularly important piece of information that we have
not utilized in this paper is that the $\taures'$ histograms could be
constructed {\em as a function of $\lobs$}, instead of lumping them
into a single histogram regardless of wavelength.  In general, if a
large number of spectra were available, this would contain more
information, and would help characterize density biases, which should
be a strong function of distance away from the source.

More specifically, however, Mesinger \& Haiman (2004) (hereafter,
MH04) recently presented a method, which takes only a crude account of
the wavelength--dependence of the opacity, using essentially only the
location of the \lya and \lyb GP troughs, to find a robust
determination of $R_S$.  Since the different hydrogen Lyman
transitions have disparate oscillator strengths, simultaneously
considering the measured absorption in two or more Lyman lines can be
an effective way to probe a sudden growth of the \lya optical depth
near the boundary of the Str\"omgren sphere.  In MH04, we applied this
technique to model the general behavior of the observed spectrum of
SDSS J1030+0524 close to the onset of the \lya and \lyb GP troughs.
We obtained a tight and robust constraint on $R_S$ (better than 5\%),
arising from the sharpness of growth of the \lya optical depth.  If
this (or any other) method can be used to independently constrain
$R_S$ to within $\sim$ 10\%, it would break the degeneracy between
small $\nf$, small $R_S$, and large $\nf$, large $R_S$ parameter
choices.  Specifically, if $R_S$ is known to within $\sim$ 10\%, a
neutral universe could be distinguished from a $\nf < 0.008$ universe
with 99\% confidence using an average of only 1 source, following the
method presented in this paper (note that an independent constraint on
the neutral fraction can be derived from the {\it size} of the HII
region, by utilizing the estimated ionizing flux of the source; Wyithe
\& Loeb 2004).  The usefulness of knowing $R_S$ can be seen from
Figure 2, where all panels have the same $R_S$ and there is a large
difference between the flux decrements arising from a neutral ({\it
left panels}) and an ionized ({\it right panels}) IGM.  This is a
highly encouraging result, and the we plan to apply both the method in
MH04 and the method presented in this paper to the current sample of
SDSS quasars.

\section{Conclusions}
\label{sec:conclude}

Through an analysis of mock quasar absorption spectra based on a
detailed cosmological hydrodynamic simulation, we have shown that it
is possible to detect the effects of the damping wing of absorption by
neutral hydrogen atoms in the IGM on top of the resonant absorption
from within the local HII region of the quasar.  We have described an
inversion method we have developed to extract an estimate of the mean
neutral fraction of hydrogen in the IGM, and of the size of the
Str\"omgren sphere around a high--redshift source.  The method is
designed to differentiate between sources embedded in an IGM with
$10^{-3}<\nf<1$, and we have found that it can distinguish among
neutral fractions in this range with only a few bright quasars.

We have explicitly incorporated into our analysis an error in the
intrinsic emission template, consisting of either an uncertainty in
its spectral power--law index, or Gaussian, uncorrelated,
pixel--to--pixel variations at each wavelength.  With both of these
errors, we find that a neutral universe can be statistically
distinguished from a $\nf = 0.008$ universe in our parameter space,
using tens of bright quasars, a sample that can be expected by the
completion of the Sloan Digital Sky Survey.  Alternatively, similar
statistical constraints can be derived from the spectra of several
hundred sources that are $\sim 100$ times fainter.  For example, the
Large-aperture Synoptic Survey Telescope (LSST) should be able to
deliver many new faint quasars that could serve as targets for
low--resolution spectroscopy (Mesinger et al. 2004, in preparation).

Furthermore, if the size of the source's \strom\ can be
independently constrained to within $\sim$ 10\% (such as with the
method presented in MH04), the analysis presented here can distinguish
between sources embedded in an IGM with $10^{-3}<\nf<1$, using a
single source.  We plan to perform such analysis on the current
sample of high--redshift sources.

The recent discovery of Gunn-Peterson troughs in the spectra of
several $z\sim 6$ quasars in the SDSS only impose the restriction of
$10^{-3}<\nf<1$ on the neutral fraction at this redshift. Further
distinguishing between the values allowed in this range, especially
between a neutral and a mostly ionized universe, would provide
invaluable new constraints that can differentiate among various
competing reionization scenarios.

\acknowledgements{ We thank an anonymous referee for a thorough
and helpful report. Zolt\'{a}n Haiman gratefully acknowledges support
by the National Science Foundation through grants AST-0307291 and
AST-0307200 and by NASA through grant HST-GO-09793.18.  Renyue Cen
thankfully acknowledges grants AST-0206299 and NAG5-13381.}

\end{document}